\documentclass{article}

\usepackage{graphicx}
\usepackage{float}
\usepackage[
singlelinecheck=false 
]{caption}

\begin{document}
\title{\bf Null geodesics and shadow of 4$D$ Einstein-Gauss-Bonnet black holes surrounded by quintessence}

\author{{Malihe Heydari-Fard$^{1}$\thanks{Electronic address: heydarifard@qom.ac.ir} and Mohaddese Heydari-Fard$^{2}$ \thanks{Electronic address: m\_heydarifard@sbu.ac.ir}}\\{\small \emph{$^{1}$ Department of Physics, The University of Qom, 3716146611, Qom, Iran}}\\{\small \emph{$^{2}$ Department of Physics, Shahid Beheshti University, G. C., Evin, Tehran 19839, Iran}}}

\maketitle

\begin{abstract}
We study the properties of the null geodesics of four-dimensional Einstein-Gauss-Bonnet black holes surrounded by quintessence matter. Due to the quintessence correction, we discuss the radial and non-radial geodesics. For the case of non-radial geodesics we obtain the effective potential, photon sphere and impact parameter associated to the null geodesics in details. We analyze the different possible orbits of massless particles (i.e. photons) such as unstable circular orbits and unbounded orbits and investigate the role of the Gauss-Bonnet coupling $\alpha$ and the quintessence parameter $q$ on the null geodesic trajectories. Moreover, we study the effects of the model parameters on the shadow cast by such black holes. The results are compared to that of Schwarzschild black hole with and without quintessence matter and four-dimensional Einstein-Gauss-Bonnet black holes without quintessence. As a physical application of null geodesics, the deflection angle is calculated and the effect of the quintessence matter on it is investigated.
\vspace{5mm}
\vspace{1mm}
\end{abstract}

\section{Introduction}
Black holes are the most fascinating prediction of Einstein's general relativity theory. The first indirect evidence for an actual black hole in nature came in 1971, from the X-ray astronomy . This astronomical X-rays source was discovered in 1965 \cite{cygnus} so called Cygnus X-1 and was  the first source extensively accepted to be a stellar-mass black hole via dynamical observations \cite{cygnus1}--\cite{cygnus2}. In recent years, observational evidence strongly indicates for the existence of black holes in two mass ranges. The LIGO detectors in USA and Virgo detectors in Italy discovered the gravitational waves produced by the merger of a binary stellar-mass black hole in 1.4 billion years ago \cite{gw}. Also, strong astronomical observations in recent years show the presence of a supermassive black hole, i.e Sgr A* at the center of Milky Way galaxy \cite{Sgr}. The supermassive black holes of $10^6-10^9$ solar masses are believed to be at the center of many galaxies in the universe. Although the black holes have been considered as possible explanation for some observed phenomena but not detected directly. Until 10 April 2019 when Event Horizon Telescope (EHT) through an international collaboration provided the first direct visual evidence for the existence of black holes. EHT revealed the shadow images of M87* \cite{eht1}--\cite{eht3} an elliptical galaxy whose supermassive black hole is 6 billion times heavier than the Sun. Therefore two astrophysical tools the study of the black hole shadow and the gravitational lensing could be used to understand the strong field regime of gravity and to test the modified gravity theories. The shadow of a Schwarzschild black hole was first discussed by Synge \cite{Synge} and further explored by Luminet in \cite{Luminet}. Bardeen then studied the shadow cast by a Kerr black hole \cite{Bardeen}. Also, the study of the black hole shadow in modified theories of gravity has attracted considerable attention in recent years \cite{shadow0}--\cite{shadow14}. For a short review on shadow of black holes see \cite{Moffat} and \cite{Herdeiro}.

The study of null geodesics around black holes is an important subject and helps us to understand the properties of black holes and the geometric structure of space-time. The exact analytical solutions of the geodesics equations help us to have the best understanding of the particles motion along the geodesics. But this is not always possible. Therefore, we can employ the numerical solutions and the analytical approximation schemes. The first study of the geodesic equations and their analytical solutions for the Schwarzschild space-time in terms of the Weierstrass elliptic functions was carried out by Hagihara in 1931 \cite{Hagihara}. Then, Darwin solved the geodesic equations using the Jacobian elliptic functions \cite{Darwin}. Moreover, the analytical solutions of the geodesic equations in four-dimensional Schwarzschild space-time \cite{Sch}, Schwarzschild-(anti) de Sitter and higher-dimensional Schwarzschild-(anti) de Sitter \cite{Sds1}--\cite{Sds2}, the Kerr space-time \cite{Kerr1}--\cite{Kerr2}, Kerr-(anti) de Sitter \cite{KerrdS1}--\cite{KerrdS2}, Kerr-Newman \cite{KN}, Reissner-Nordstrom \cite{RN1}--\cite{RN3}, Reissner-Nordstrom-(anti) de Sitter \cite{RNdS} and Myers-Perry space-times \cite{Myers1}--\cite{Myers2} have been investigated. The study of geodesic motion in the context of modified theories of gravity have been also extensively studied. The geodesic motion of test particles in $f(R)$ modified gravity have been studied in \cite{fR}. In the context of the brane world scenario, the null geodesics have been considered in \cite{brane1}--\cite{brane2}. The geodesic analysis around black holes in Horava-Lifshitz gravity, Einstein-Maxwell-dilaton gravity and conformal weyl gravity have been discussed in \cite{Horava1}--\cite{Weyl2}, respectively. The null geodesics of Born-Infeld black holes have been considered in \cite{BI1}--\cite{BI4}. In the space-time of the Schwarzschild and Kerr black holes pierced by a cosmic string, test particles motion has been studied in \cite{C1}--\cite{C2}. In \cite{N1}--\cite{N2}, the null geodesics of noncommutative geometry have been considered. For study of time-like and null-like geodesics around the black holes surrounded by quintessence matter and around hairy black holes, see \cite{Qu1}--\cite{Qu5} and \cite{Hair1}--\cite{Hair2}, respectively. Also, in the background of wormhole geometries the geodesic analysis have been considered in \cite{Wh1}--\cite{Wh3}.

In the past decades, the modified theories of gravity have been taken considerable interest to address the theoretical and observational issues that can not be explained in the framework of Einstein's theory of general relativity. A well-known example in this regard is the Gauss-Bonnet (GB) gravitational theory which appears in the low energy limit of the string theory \cite{string1}--\cite{string2}. As we know, in four-dimensional space-time, the integral of the GB term is a topological invariant and does not contribute to the gravitational field equations, unless a scalar field is coupled to the GB term by a regular coupling function. In this context, the black hole solutions, with various coupling functions, have been well studied since 1990s \cite{st1}--\cite{st6}.

However, recently Glavan and Lin proposed a novel 4-dimensional Einstein-Gauss-Bonnet (4$D$ EGB) gravity \cite{Glavan} that by re-scaling the GB coupling as $\alpha\rightarrow\frac{\alpha}{(D-4)}$ and taking the limit $D\rightarrow4$, circumvent the conditions of Lovelock’s theorem. In such a way, the GB term can make a non-trivial contribution to the gravitational field equations. This theory preserves the number of degrees of freedom and avoids the Ostrogradsky instability. But, it is found that taking the limit $D\rightarrow4$ may not be consistent and the vacua of the model are not also well-defined \cite{c1}--\cite{c7}. To solve these problems, some alternative approaches such as introducing a counter term into the action and then take $D\rightarrow4$ limit \cite{s1}--\cite{s2}, compactification of $D$-dimensional EGB gravity \cite{s3}--\cite{s4} and breaking the temporal diffeomorphism invariance \cite{s5} have been suggested to obtain a consistent EGB gravity. It is worth mentioning that these consistent theories share the black hole solutions with the original 4$D$ EGB gravity that attracted a great deal of attention. For instance, the black hole solutions \cite{charge}--\cite{bardeen}, quasinormal modes and stability analysis of black holes \cite{QNM1}--\cite{stability3}, thermodynamics and weak cosmic censorship  \cite{th1}--\cite{th4}, thin accretion disks \cite{disk1}--\cite{disk2}, test particle motion and black hole shadow \cite{isco}--\cite{GB5}, gravitational lensing \cite{lensing1}--\cite{lensing4} and also some cosmological aspects of 4$D$ EGB gravity \cite{cosmology1}--\cite{cosmology6} have been extensively studied. In this paper, we consider 4$D$ EGB black holes surrounded by quintessence matter \cite{QEGB} to study the null geodesics around them and investigate the effects of the model parameters on these trajectories.

The paper is structured as follows. In section 2, we present a brief review of 4$D$ EGB black holes surrounded by quintessence matter and some of their properties. A detailed analysis of radial and non-radial null geodesics around these black holes is done in section 3. In section 4, the shadow radius of quintessence 4$D$ EGB black holes is calculated. Then, as an application of null geodesics, we study the bending of light in the space-time of such black holes in section 5 and finally we summarize the conclusions in section 6.

\section{4$D$ EGB black holes with quintessence matter}
In this section we introduce 4$D$ EGB black hole solutions surrounded by quintessence matter obtained in \cite{QEGB}. The line element of this solution has the following form
\begin{equation}
ds^2=-f(r)dt^2+\frac{dr^2}{f(r)}+r^2\left(d\theta^2+\sin ^2\theta d\varphi^2\right),
\label{1}
\end{equation}
where
\begin{equation}
f_{\pm}(r)=1+\frac{r^2}{2\alpha}\left[1\pm\sqrt{1+\frac{8\alpha M}{r^3}+\frac{8\alpha q}{r^{3\omega_{q}+3}}}\right],
\label{2}
\end{equation}
that $M$ is the black hole mass, $\alpha$ is the GB coupling constant and $\omega_q$ and $q$ are the state parameter of the quintessence matter and the normalization factor (quintessence parameter), respectively. It is clear that in the case of $q=0$, the above solution reduces to the 4$D$ EGB black holes obtained by Glavan and Lin \cite{Glavan}. In the limit of $\alpha\rightarrow 0$, the negative branch corresponds to the Schwarzschild solution surrounded by quintessence matter, that was initially obtained by Kiselev in \cite{Kiselev}. Thus, in this paper we consider the negative branch of the above solution. Also, the case of $q = 0$ and $\alpha\rightarrow 0$ represents the ordinary Schwarzschild solution. Moreover, by choosing the different values of the state parameter $\omega_q$, the different black hole solutions can be obtained. For instance, in the case of no quintessence $\omega_q=0$ the above metric reduces to 4$D$ EGB black holes \cite{Glavan}, and for $\omega_q=-1$ and $q=\frac{\Lambda}{6}$ it reduces to 4$D$ EGB de Sitter solution\cite{charge}. When $\omega_q=\frac{1}{3}$ and $q=-\frac{Q^2}{2}$, the solution (2) corresponds to 4$D$ charged EGB black holes obtained in \cite{charge}.

The behaviour of the $f(r)$ function and thus the horizons of black hole, strongly depends on the values of $\omega_q$, $q$ and also the GB coupling constant $\alpha$. The state parameter $\omega_q$ must be in the range $-1 <  w_q < -\frac{1}{3}$, and the energy density of the quintessence matter is as follows \cite{QEGB}
\begin{equation}
\rho_q=-6 w_q q \frac{1}{r^{3(1+w_q)}},
\label{4}
\end{equation}
as is clear, in order to have a positive energy density the $q$ parameter must be positive. In Figure 1 we have plotted the behaviour of the metric function $f(r)$ for different values of $w_q$. For the case of $\alpha > 0$, it is clear that for central values of $w_q$ in the interval $w_q \in (-1,-\frac{1}{3})$ there are three horizons, but when approaching to the boundary value in the `right' there is two horizon and approaching to the `left' there is one horizon. For further study we refer the interested reader to \cite{QEGB}.

In the present work, we are interest to consider $\omega_q = -0.35$ and thus the function of $f(r)$ in the equation (\ref{2}) takes the following form
\begin{equation}
f(r) = 1+ \frac{r^2}{2 \alpha } \left[1-\sqrt{1+\frac{8 \alpha  M}{r^3}+\frac{8 \alpha  q}{ r^{1.95}}}\right].
\label{5}
\end{equation}
The behaviour of the above metric function $f(r)$ for $\alpha>0$ and $\alpha<0$ with different values of the $q$ parameter has plotted in Figure 2. The figure shows that for positive values of $\alpha$, the function of $f(r)$ has two real roots that are identified as the cosmological quintessential horizon $r_{-}$ and the event horizon $r_+$, while for negative values of $\alpha$ there is only one horizon at $r=r_+$.

As we mentioned before, the case of $q=0$ corresponds to 4$D$ EGB black holes without quintessence. In this case, the inner and outer event horizons are given by
\begin{equation}
r_{\pm}={M\pm\sqrt{M^2-\alpha }},
\label{6}
\end{equation}
where for $\alpha>0$ there are two horizons while for $\alpha<0$ there is only one horizon. However, in the case of $f(r)$ function in equation (\ref{5}), we have not an explicit expression for the event horizon and thus we have numerically displayed the behaviour of $r_{+}$ and $r_-$ as a function of the GB coupling in Figure 3. It is easy to see that with increasing $\alpha$ the event horizon $r_+$ decreases. This is due to the fact that for $\alpha >0$, the GB term play the role of dark energy, counteract gravity and thus the event horizon radius takes smaller values. Also, we see that by increasing $q$, the event horizon $r_+$ increases while, the quintessential horizon $r_-$ decreases. Therefore, the presence of quintessence matter around 4$D$ EGB black holes increases the event horizon. The numerical results have also summarized in Table 1. For $q=0$ the results of Table 1 are the same as the values of Table 1 in ref \cite{GB2}.

\begin{figure}[H]
\centering
\includegraphics[width=3.0in]{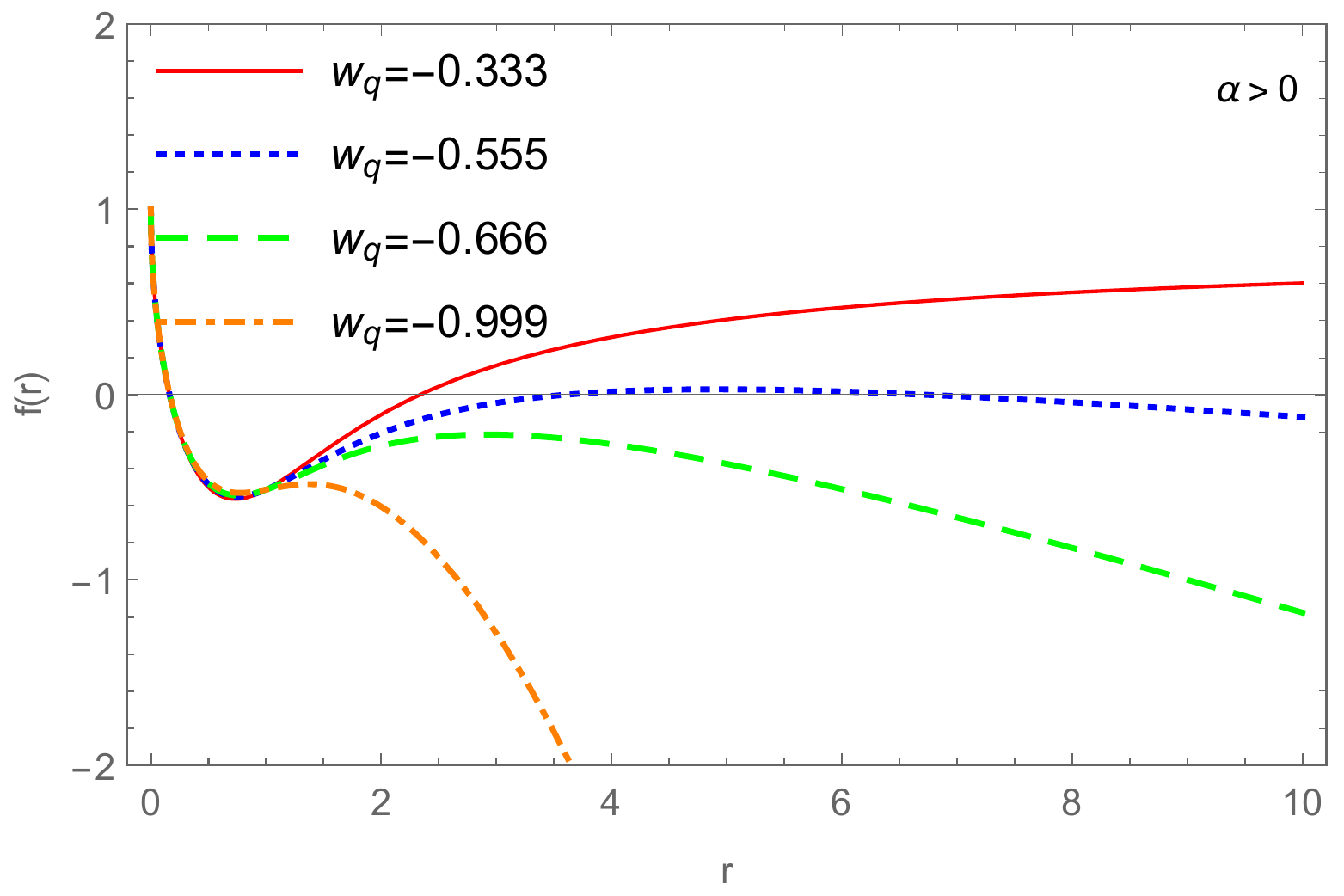}
\includegraphics[width=3.0in]{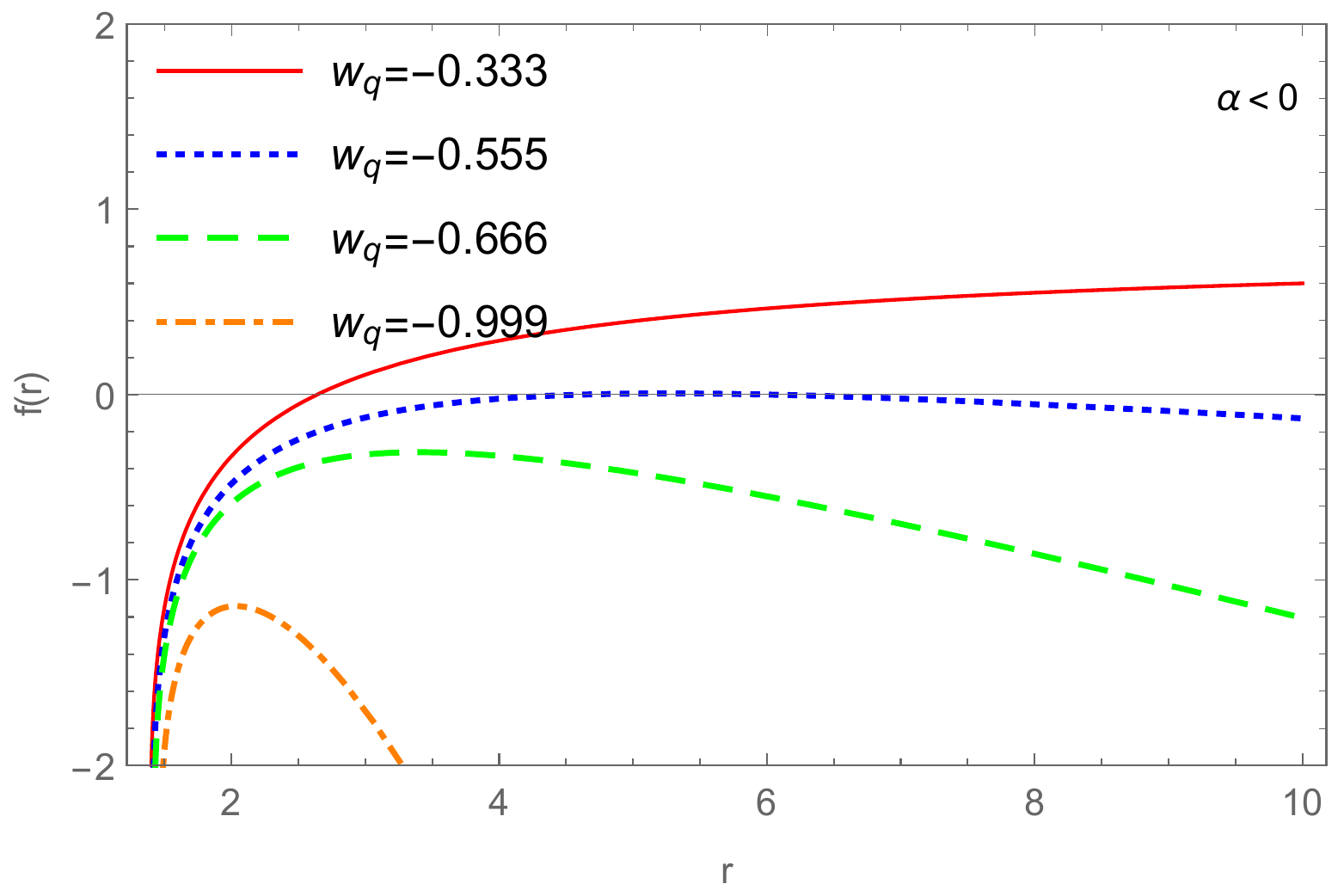}
\caption{The metric coefficient $f(r)$ as a function of the radial coordinate $r$ for different values of $w_q$. Left panel: for positive GB coupling $\alpha = 0.3$. Right panel: for negative GB coupling $\alpha = -0.3$ with $q = 0.1$ and $M = 1$.}
\label{figure-2}
\end{figure}

\begin{figure}[H]
\centering
\includegraphics[width=3.0in]{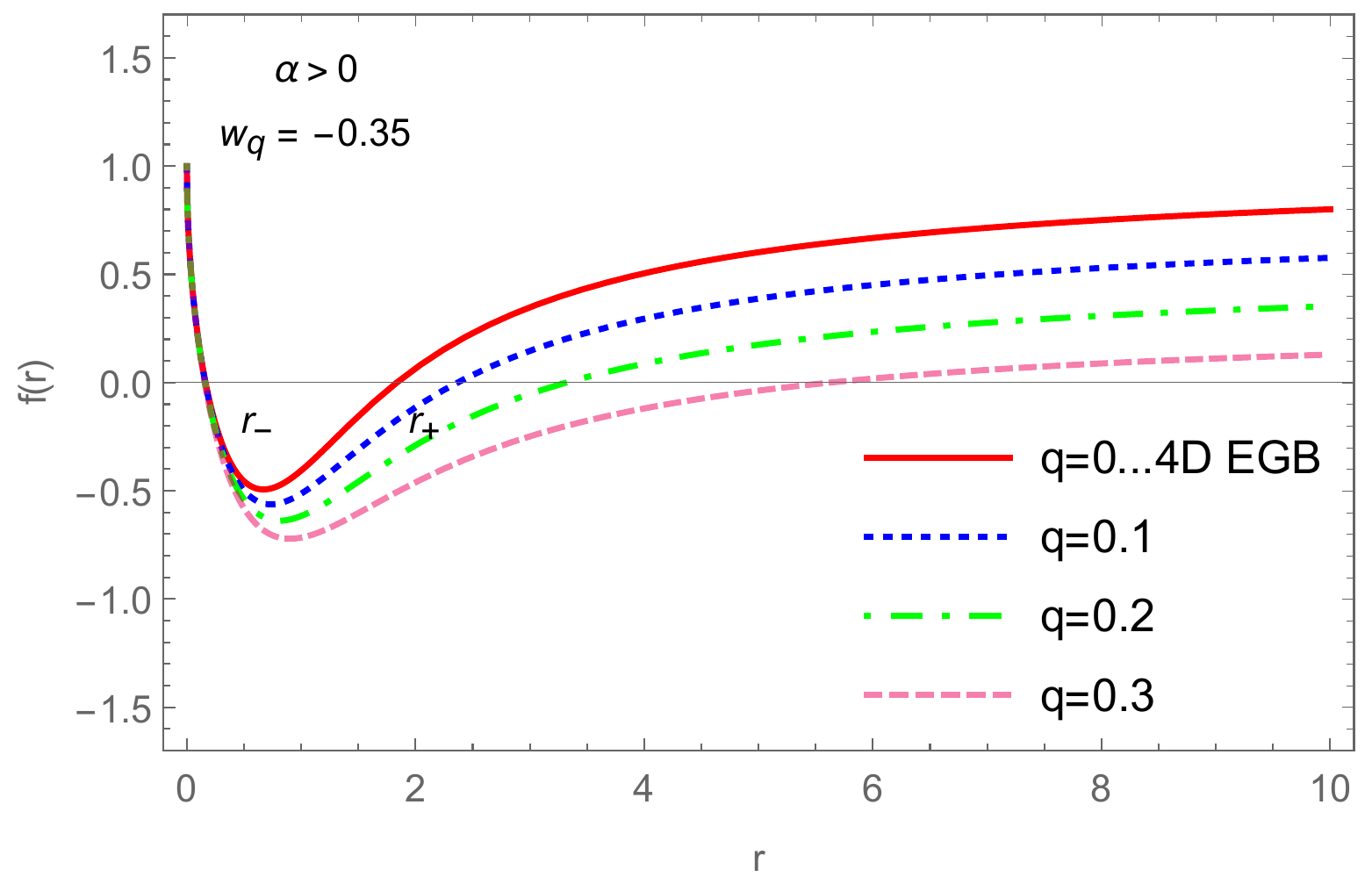}
\includegraphics[width=3.0in]{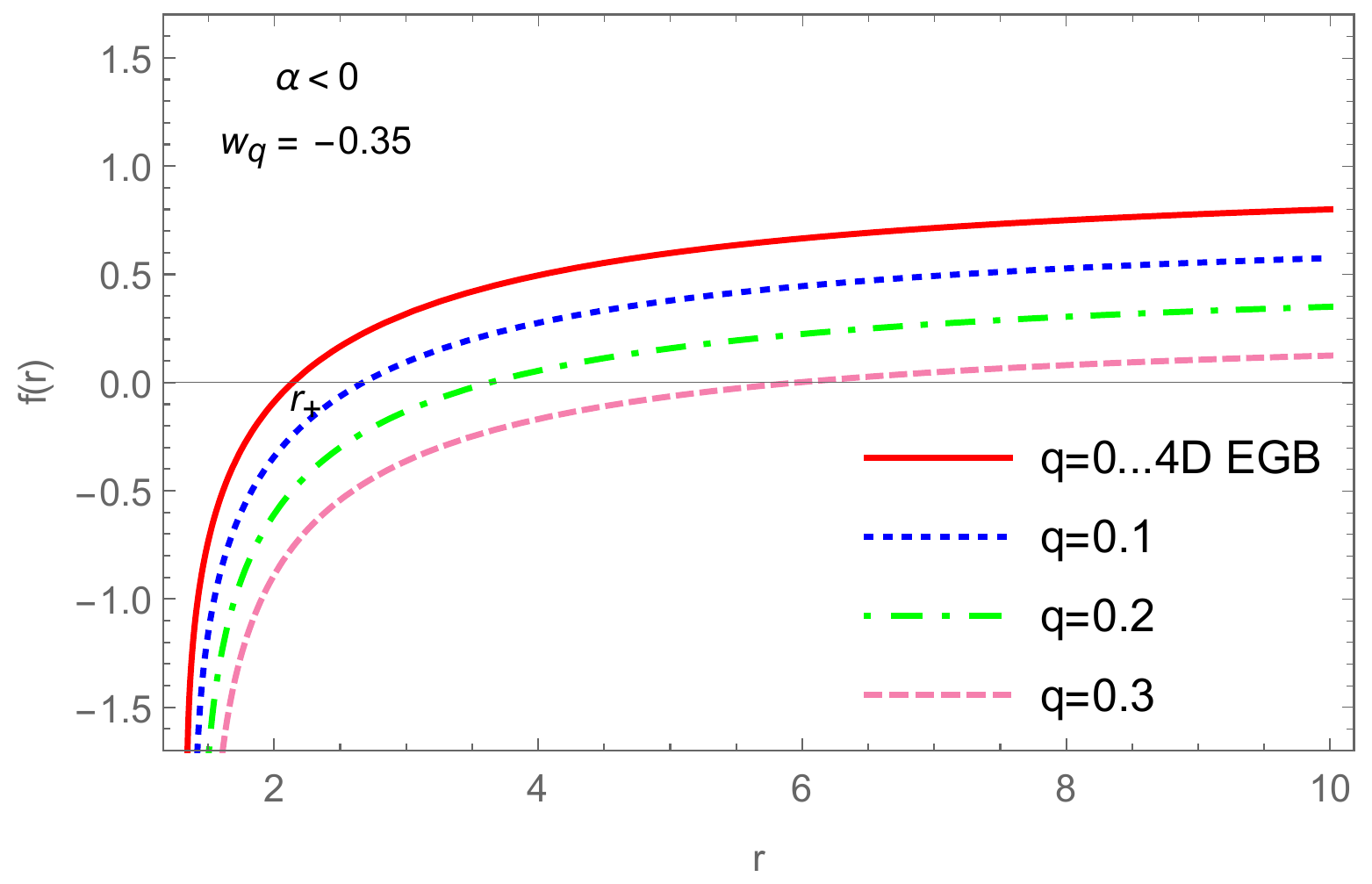}
\caption{The metric coefficient $f(r)$ as a function of the radial coordinate $r$ for different values of $q$. Left panel: for positive GB coupling $\alpha = 0.3$. Right panel: for negative GB coupling $\alpha = -0.3$ and $M = 1$.}
\label{figure-1}
\end{figure}

\begin{figure}[H]
\centering
\includegraphics[width=3.0in]{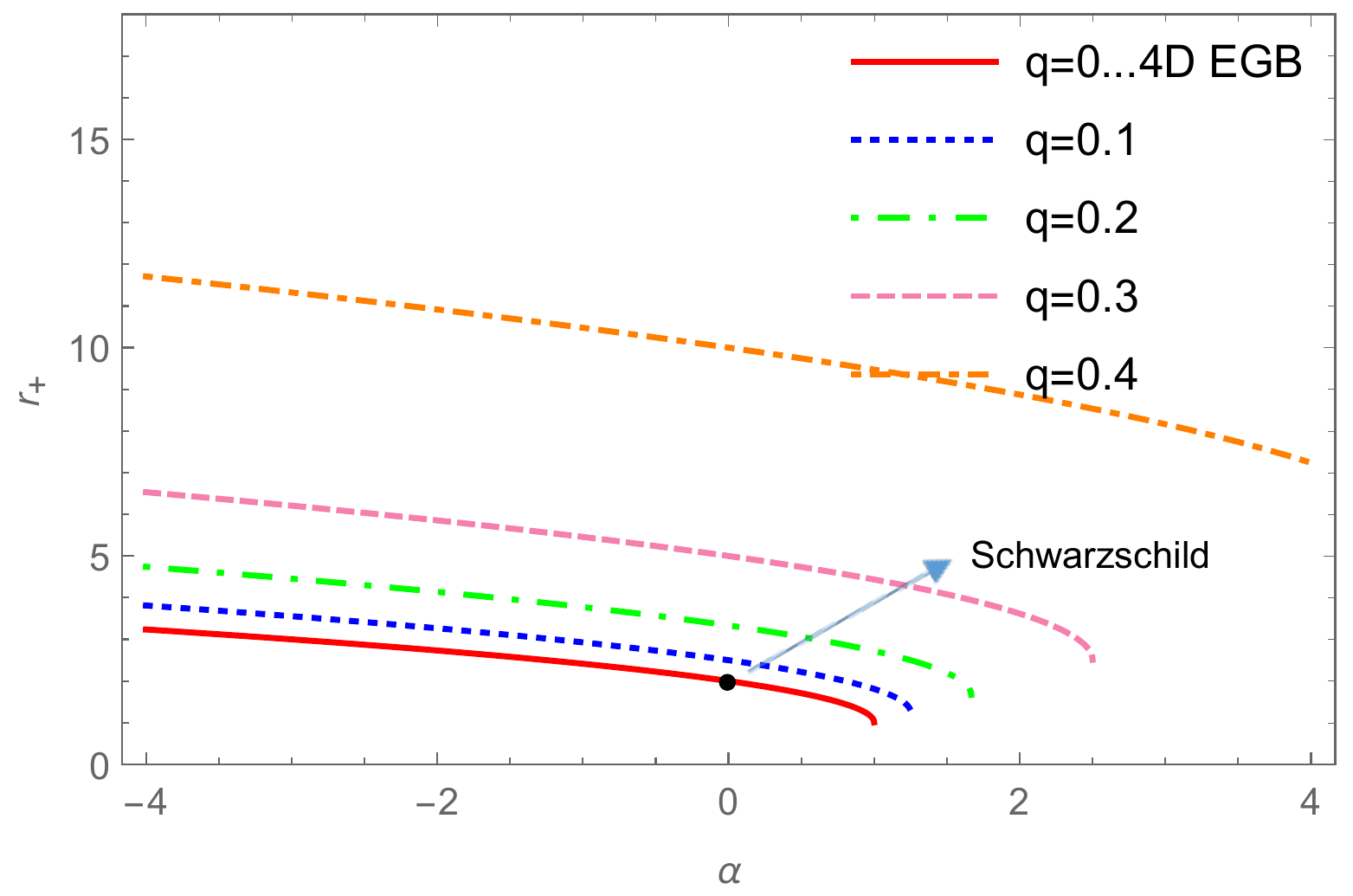}
\includegraphics[width=3.0in]{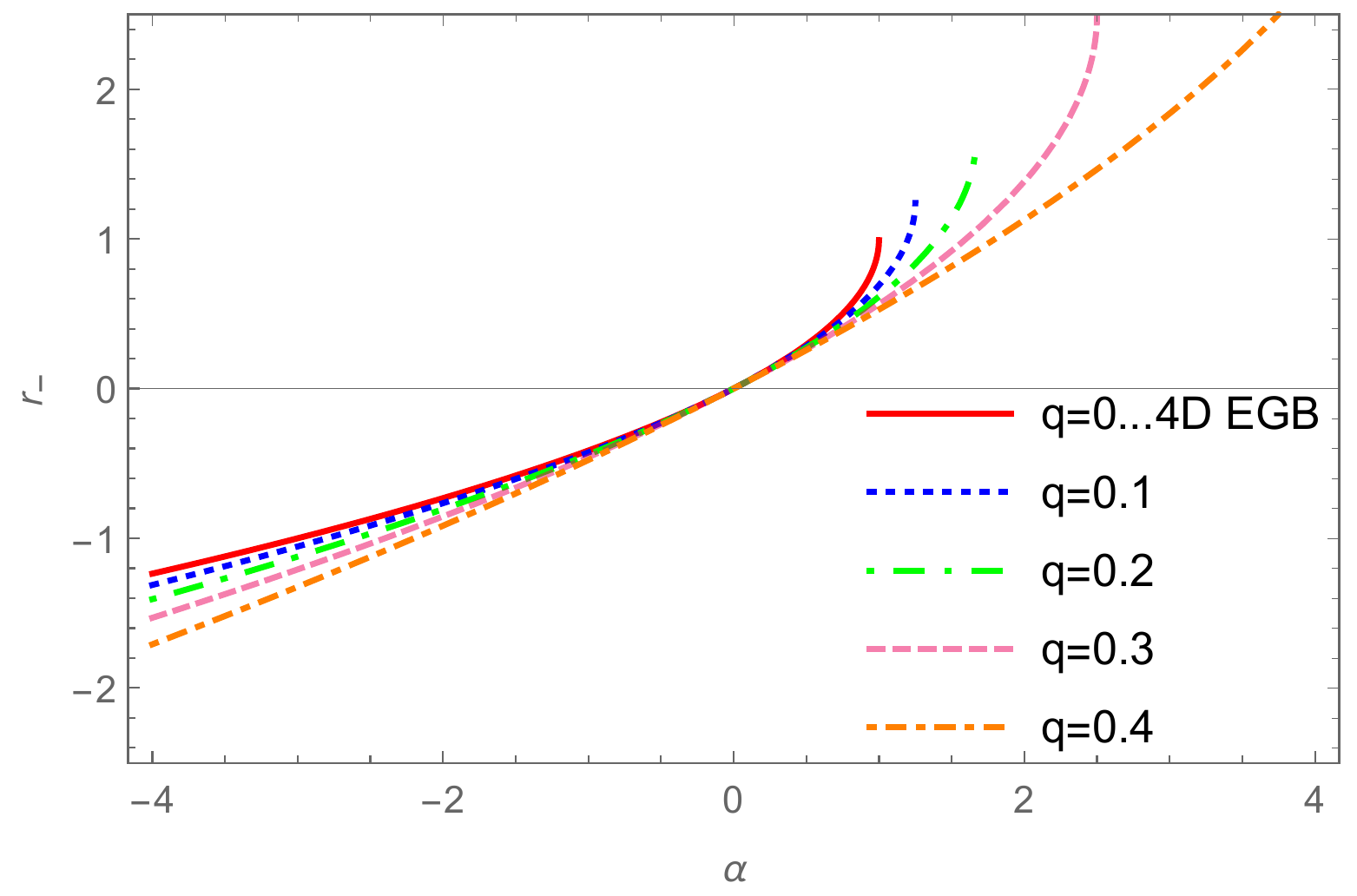}
\caption{ Left panel: the behavior of the event horizon $r_{+}$ as a function of GB coupling. On the red curve, the point of $\alpha = q= 0$ corresponds to the Schwarzschild event horizon at $r_+ = 2M$. Right panel: the cosmological quintessential horizon $r_{-}$ as a function of GB coupling for $M = 1$.}
\label{horizon}
\end{figure}

\begin{table}[H]
\centering
\caption{\footnotesize The quintessential cosmological horizon $r_{-}$, the event horizon $r_{+}$, the radius of the photon sphere $r_c$ and the impact parameter $b_c$ of the photon sphere for different values of $\alpha$ and $q$ with $M = 1$.}
\begin{tabular}{l l l l l l l}
\hline\hline
$q$&$\alpha$&$r_{-}/M$& $r_{+}/M$&$r_{c}/M$&$b_{c}$\\ [0.5ex]
\hline\hline\\
{0}
     &0.7&0.45227&1.54772&2.61959&4.87959\\
    &0.5& 0.29289&1.70711&2.74656&4.98198\\
     &0.3&0.16334&1.83666&2.85656&5.07351\\
     &0.1&0.05132&1.94868&2.95453&5.15692\\
     &0&0&2              &3      &5.19615\\
    &-1.1&NH&    2.44914 &3.40373&5.55557\\
    & -3.3&NH&   3.07364&3.95844&6.07084\\
     &-5.5&NH&   3.54951 &4.36744&6.64004\\
     &-7.7&NH&3.94958&4.70134&6.78150\\
\hline\hline\\
{0.1}
     &0.7&0.42197&2.10407&3.38368&7.01918\\
     &0.5&0.28175&2.24472&3.51361&7.15518\\
     &0.3&0.16027&2.36768&3.63108&7.28079\\
     &0.1&0.05104&2.47833&3.73897&7.39808\\
     &0&0        &2.53005& 3.78995 &7.45401\\
     &-1.1&    NH&3.00061&4.26285&7.98741\\
     &-3.3&    NH&3.68224&4.95629&8.80054\\
     &-5.5    &NH&4.21107&5.49176&9.44392\\
     &-7.7&NH&4.65903&5.94183&9.99154\\
\hline\hline
\end{tabular}
\end{table}

\section{Null geodesics of quintessence 4$D$ EGB black holes }
The trajectory of photons in quintessence background of 4$D$ EGB black holes describe by the null geodesics. The geodesic equations for this space-time can be obtained from the following Lagrangian
\begin{equation}
{\cal L}=\frac{1}{2}\left(-f(r) \dot{t}^2+\frac{1}{f(r)} \dot{r}^2 + r^2 \dot{\theta}^2+ r^2 \sin^2{\theta}\dot{\varphi}^2\right),
\label{7}
\end{equation}
where a dot denotes differentiation with respect to the proper time $\tau$. Using the Euler-Lagrange equations for $t$ and $\varphi$ coordinates we have
\begin{equation}
\dot{t} =\frac{E}{f(r)},
\label{8}
\end{equation}
\begin{equation}
\dot{\varphi} = \frac{L}{r^2 \sin^2{\theta}},
\label{9}
\end{equation}
where $E$ and $L$ are two constants of motion, namely the energy and angular momentum of test particles. We consider the motion in the equatorial plane, $\theta=\frac{\pi}{2}$, and thus we have $\dot{\theta}=\ddot{\theta}=0$. By substituting $\dot{t}$ and $\dot{\varphi}$ in the equation (\ref{7}) the lagrangian takes the following form
\begin{equation}
\dot{r}^2+f(r)\left(\frac{L^2}{r^2}+h\right)=E^2,
\label{10}
\end{equation}
where $2{\cal L}=h$ and $h=1$ and $h=0$ correspond to the time-like and null-like geodesics, respectively. Comparing the above equations to the relation $\dot{r}^2+V_{\rm eff}=E^2$, one can find the effective potential as
\begin{equation}
V_{\rm eff}=\left(\frac{L^2}{r^2}+h\right)f(r).
\label{11}
\end{equation}

Since in the present work we focus on the null geodesics with $h=0$, the effective potential is given by
\begin{equation}
V_{\rm eff}=L^2\frac{f(r)}{r^2}.
\label{12}
\end{equation}
To describe the photon trajectories, one can find the following equation of motion by eliminating the $\tau$ parameter from equations (\ref{9}) and (\ref{10})
\begin{equation}
\left(\frac{dr}{d\varphi}\right)^2=\frac{r^4}{b^2}-r^2f(r),
\label{13}
\end{equation}
where the impact parameter is defined as $b\equiv\frac{L}{E}$.

In the next section we analyze both the radial and non-radial null geodesics and investigate the role of the GB coupling constant $\alpha$ and quintessence parameter $q$ on them.

\subsection{Radial null geodesics $(L=0)$}
The radial geodesics correspond to the motion of particles with vanishing angular momentum, $L=0$. So, the effective potential for radial null geodesics is zero, $V_{\rm eff}=0$, and the equations of motion for $\dot{r}$ and $\dot{t}$ are now given by
\begin{equation}
\dot{r}=\frac{dr}{d\tau}=\pm E,
\label{14}
\end{equation}
\begin{equation}
\dot{t}=\frac{dt}{d\tau}=\frac{E}{f(r)}.
\label{15}
\end{equation}
From the above equations we have
\begin{equation}
\frac{dt}{dr}=\pm\frac{1}{f(r)}=\pm\frac{1}{ 1+ \frac{r^2}{2 \alpha } \left[1-\sqrt{1+\frac{8 \alpha  M}{r^3}+\frac{8 \alpha  q}{ r^{1.95}}}\right]}.
\label{16}
\end{equation}
Also, the differential equation for the proper time is given by
\begin{equation}
\frac{d\tau}{dr}=\pm\frac{1}{E},
\label{17}
\end{equation}
which has a solution as
\begin{equation}
\tau=\pm\frac{r}{E}+\rm const_{\pm}.
\label{18}
\end{equation}
However, the equation (\ref{16}) does not have an analytical solution, and thus we need to solve it numerically. We have plotted the behaviour of the coordinate time $t$ and also the proper time $\tau$ in Figure 4. Note that we study the ingoing radial geodesics and thus consider the minus sign in these equations. As can be seen, when particles approach the event horizon, $r\rightarrow r_{+}$, we have $t\rightarrow\infty$, while it takes a finite period of the proper time $\tau$, which is similar to the case of the Schwarzschild black hole \cite{Sch} and the Schwarzschild black hole surrounded by quintessence \cite{Qu2}.

\begin{figure}[H]
\centering
\includegraphics[width=3.0in]{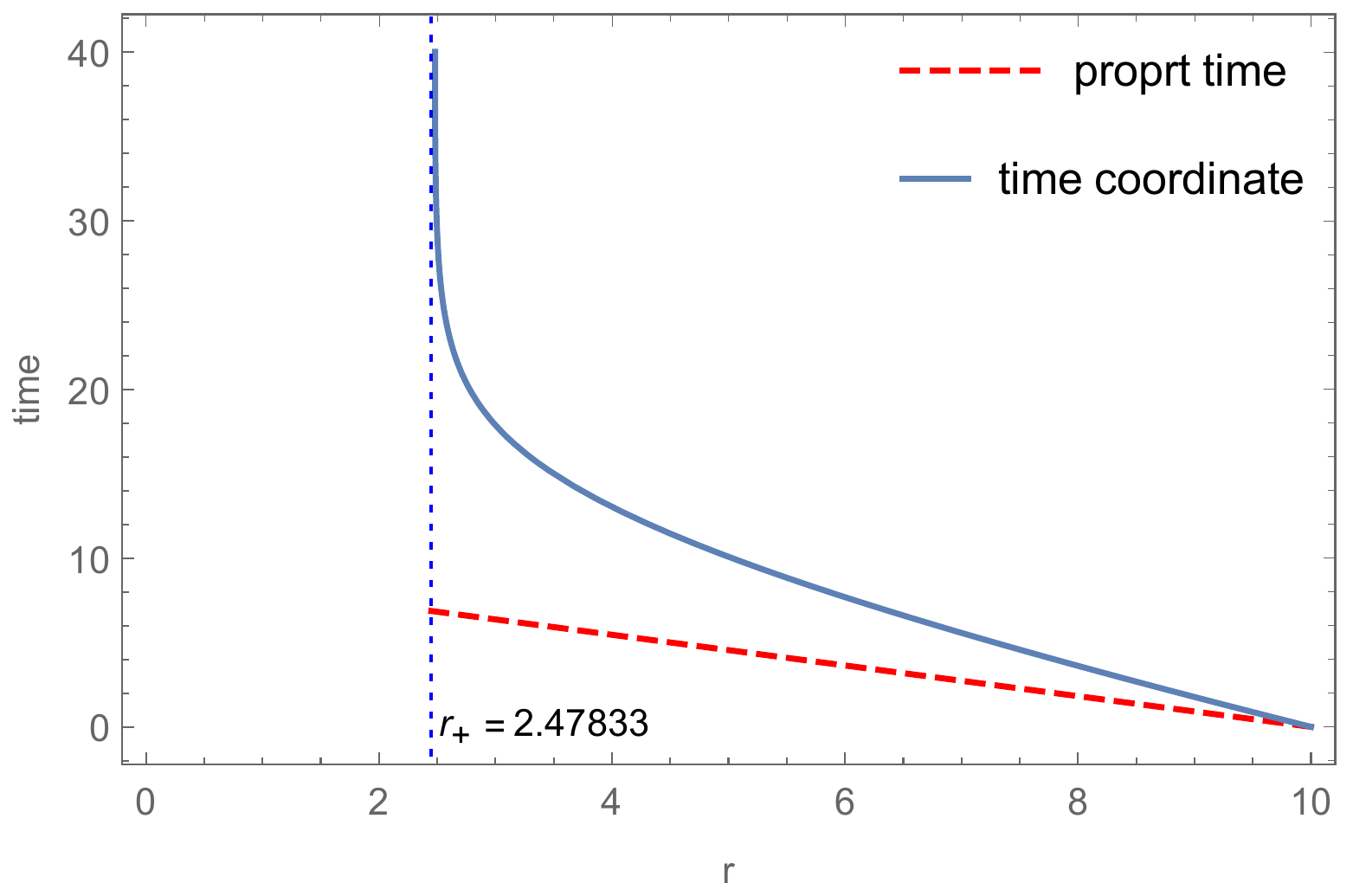}
\caption{The coordinate time $t$ and the proper time $\tau$ as a function of the radius $r$ for quintessence 4$D$ EGB black holes with $\alpha = q
 = 0.1$ and $M = 1$.}
\label{time}
\end{figure}

\subsection{Null geodesics with angular momentum ($L\neq0$)}
Now, we are going to study null geodesics with non-zero angular momentum.
\subsubsection{Effective potential}

In the case of non-radial geodesics the effective potential is given by
\begin{equation}
V_{\rm eff} =\frac{L^2}{r^2}f(r)=\frac{L^2}{r^2}\left[1+ \frac{r^2}{2 \alpha } \left(1-\sqrt{1+\frac{8 \alpha  M}{r^3}+\frac{8 \alpha  q}{ r^{1.95}}}\right)\right].
\label{19}
\end{equation}
The effective potential of massless particles in the space-time of 4$D$ EGB black holes surrounded by quintessence has plotted in Figure 5. In the left panel of the figure, we have shown the effective potential for different values of $q$ parameter with $\alpha = 0.3$ and for comparison we have also plotted it for 4$D$ EGB black holes without quintessence, $q=0$. It is easy to see that the potential for 4$D$ EGB black holes is higher than that for quintessence 4$D$ EGB black holes; namely the effect of the quintessence parameter is to decreases the effective potential. The effect of the GB coupling $\alpha$ on the effective potential has been shown in the right panel of the figure. We see that by increasing the GB coupling the effective potential increases, too. For different values of $L$, the effective potential is plotted in Figure 6.  As one expected, the effective potential is larger for larger values of $L$.

As we know, the motion of particles under the effective potential $V_{\rm eff}(r)$ depends on their energy levels. So, according to the Figure 7, by considering the photons coming from infinity and move radially inward direction, the different motions of particle, can be described as follows:

\begin{enumerate}
 \item {\bf Scattering orbits}: In region I, when the photons with $b > b_{c}$ start motion from $r > r_0$, approach to the turning point at $r = r_0$ and thus pushed away to the large $r$, again.
      \item  {\bf Falling orbits}: Another possibility in region I is the photons with $b > b_c$ starts in $r_{+} < r < r_{1}$. In such a case, photons cross the event horizon $r_{+}$ and fall into the black hole.
          \item {\bf Circular orbits}: For $b=b_{c}$, we have $\dot{r}=0$ at $r=r_c$. As is clear from the shape of the potential, if photons start at $r>r_{c}$, they will have unstable circular motion at $r=r_{c}$. The unstable circular orbits at $r=r_c$ known as the photon sphere.
   \item  {\bf Falling orbits}: In region III, the photon with $b < b_c$ that coming from infinity will continue its motion inward and finally enter the inside of the black hole.
\end{enumerate}

\begin{figure}[H]
\centering
\includegraphics[width=3.0in]{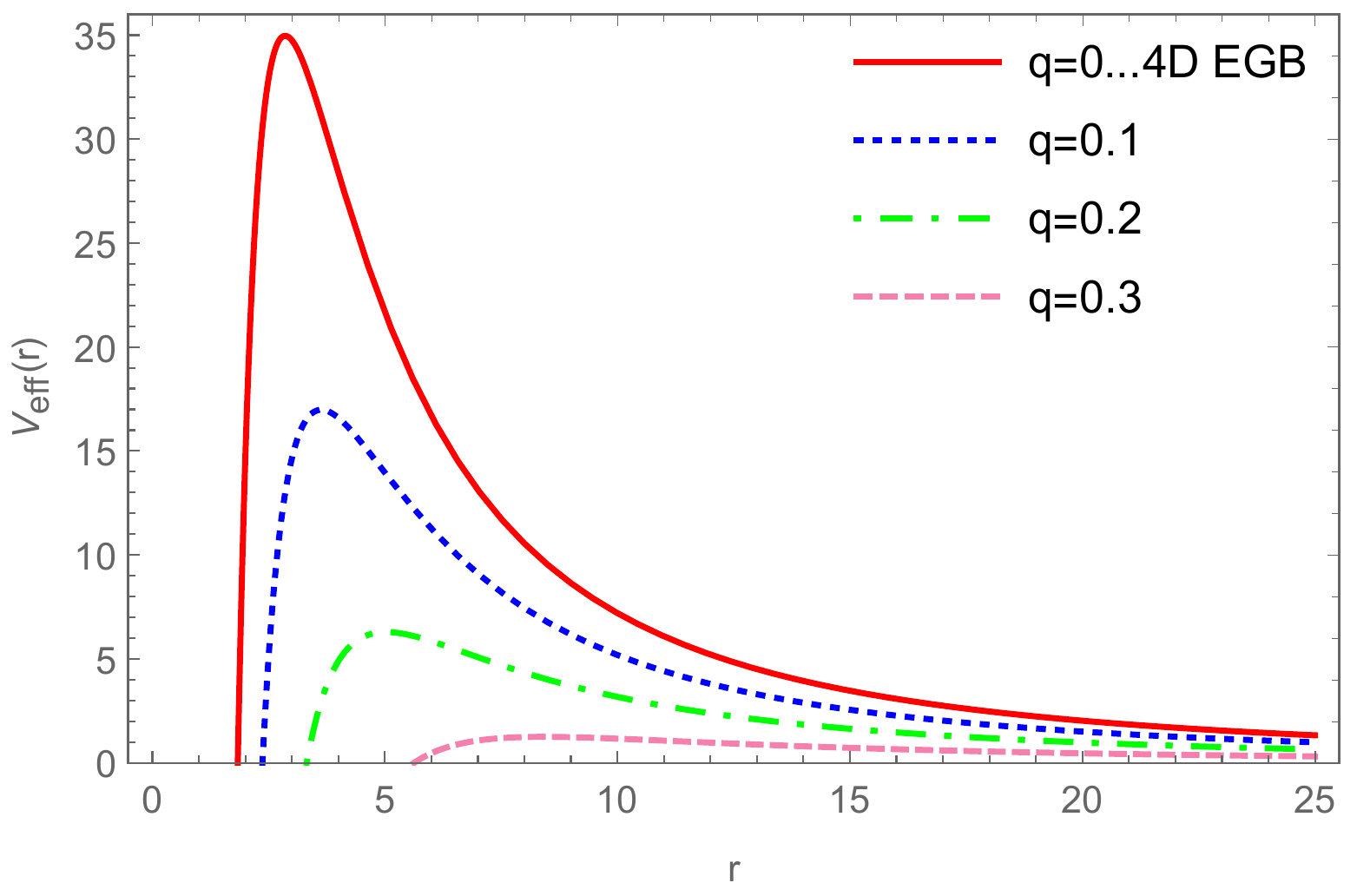}
\includegraphics[width=3.0in]{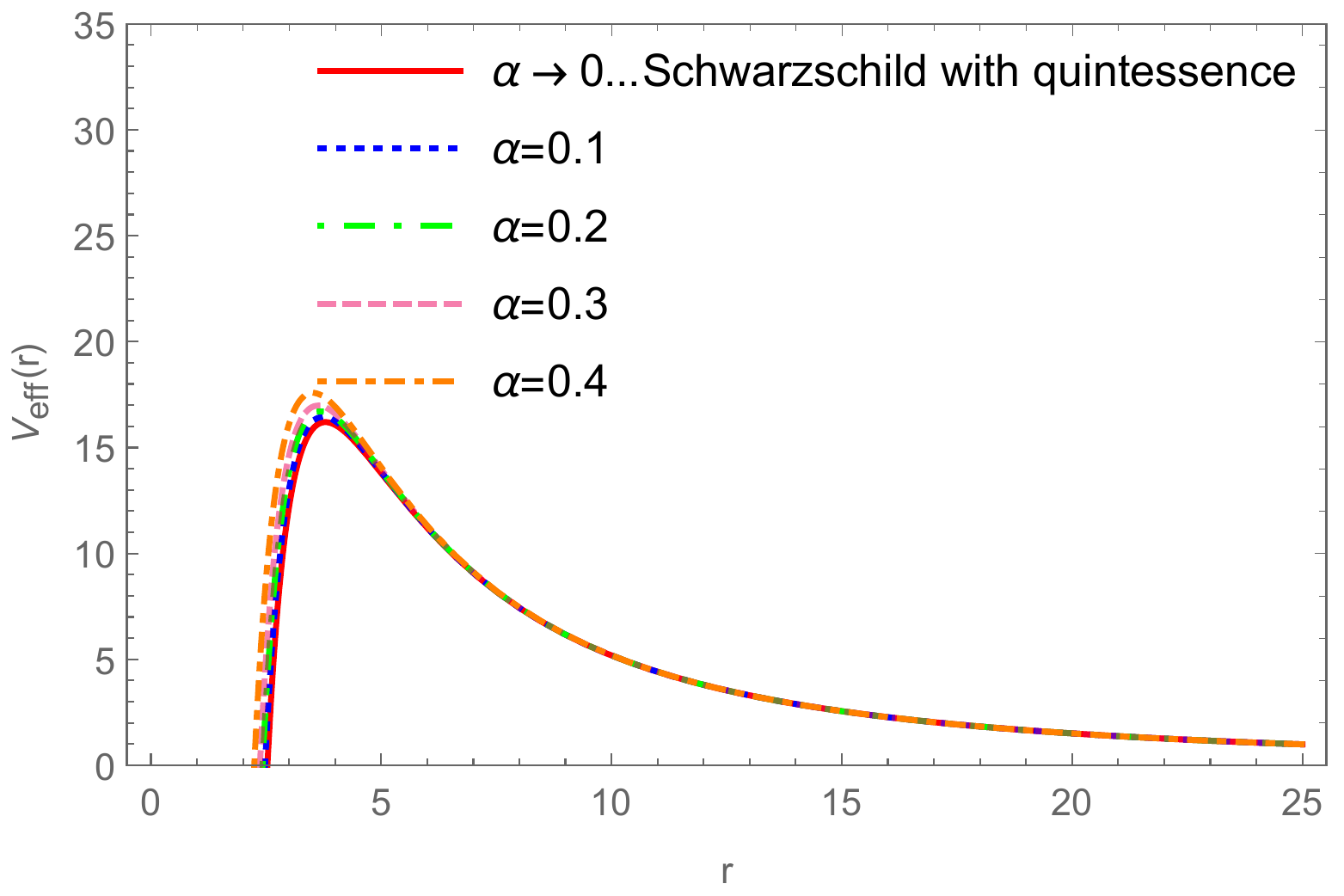}
\caption{Left panel: The effective potential as a function of $r$ for $\alpha = 0.3$ and different values of $q$. Right panel: The effective potential as a function of $r$ for $q = 0.1$ and different values of GB coupling $\alpha$, with $L = 30$ and $M = 1$. The solid curves correspond to 4$D$ EGB black holes without quintessence matter ($q=0$) and the Schwarzschild solution surrounded by quintessence ($\alpha=0$) in the left and right panels, respectively.}
\label{5-potential}
\end{figure}

\begin{figure}[H]
\centering
\includegraphics[width=3.0in]{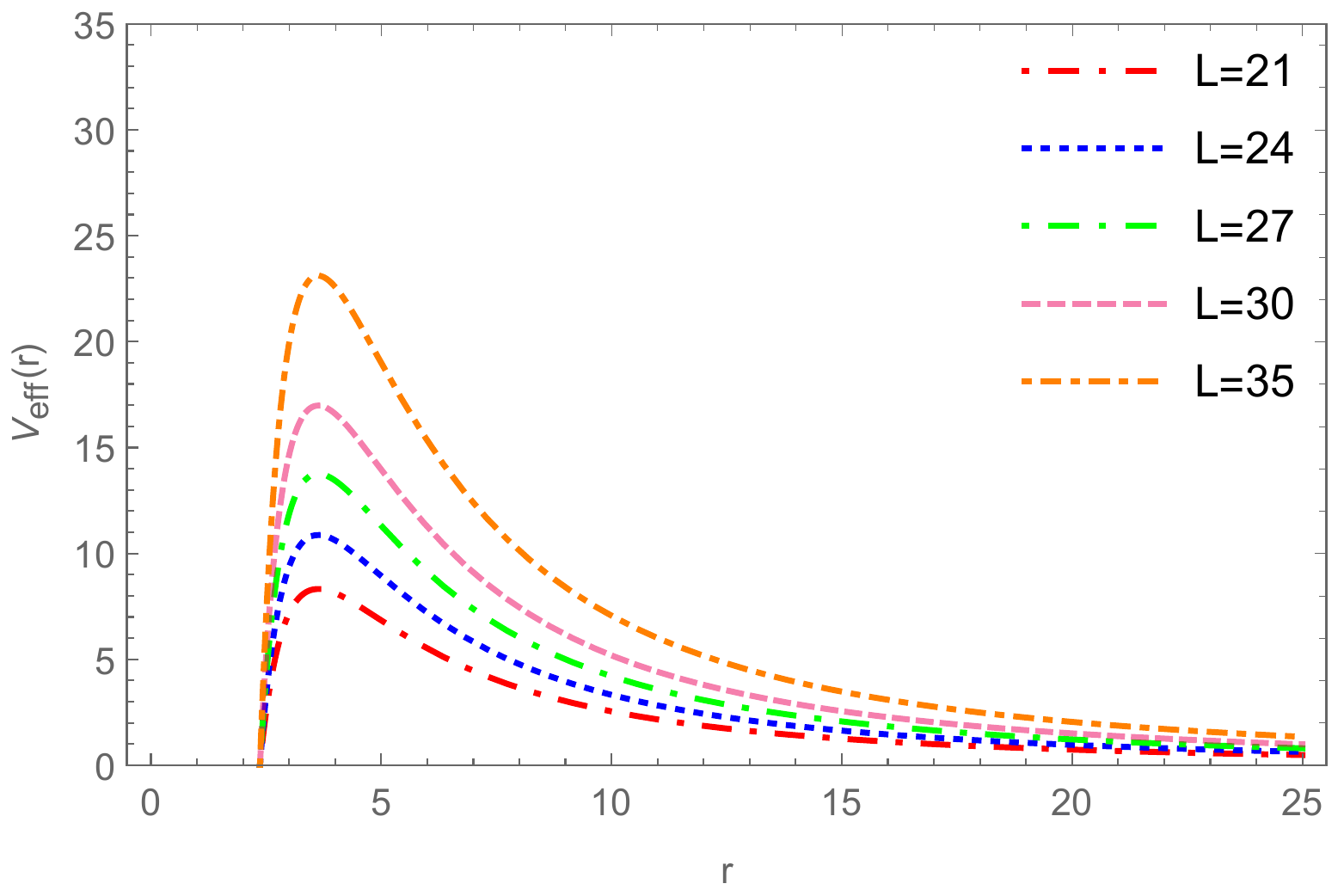}
\caption{The effective potential as a function of the radial coordinate $r$ for different values of the angular momentum $L$, with $\alpha = 0.3$, $q = 0.1$ and $M = 1$.}
\label{6-ptential}
\end{figure}

\begin{figure}[H]
\centering
\includegraphics[width=3.0in]{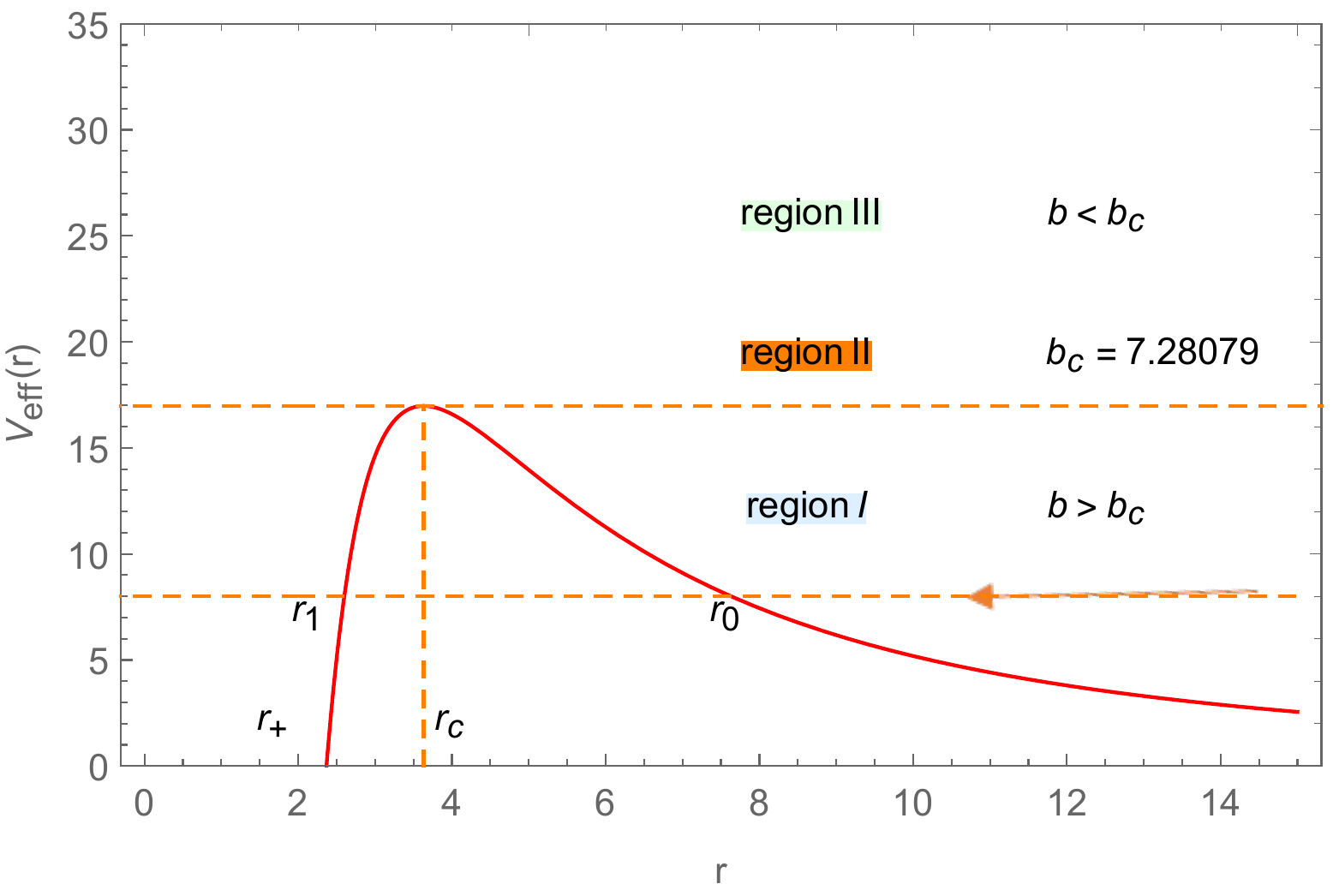}
\caption{The effective potential for quintessence 4$D$ EGB black holes for $\alpha = 0.3$, $q =0.1$ and $L = 30$ with $M = 1$. The different kinds of motion are described with different values of the impact parameter $b$.}
\label{7-potential}
\end{figure}

\subsubsection{Circular orbits}
As we mentioned in the previous section, the circular orbits occur at $r=r_{c}$. In this point
\begin{equation}
V_{\rm eff}=E_c^2,
\label{20}
\end{equation}
and thus $\dot{r}=0$ and the photon circle the black hole at a constant radius $r=r_c$. From the above equation we have
\begin{equation}
\frac{dV_{\rm eff}}{dr}=0,
\label{21}
\end{equation}
which using the equation (\ref{12}) leads to the following equation
\begin{equation}
r\frac{df(r)}{dr}-2f(r)=0.
\label{22}
\end{equation}
By substituting the $f(r)$ function from equation (\ref{2}), one can find the following equation for the photon radius
\begin{equation}
6 q (1+w)-2 r^{3 w} \left(-3 M+r \sqrt{1+\frac{8 \alpha  M}{r^3}+\frac{8 \alpha q}{r^{3(1+w)}}}\right)=0.
\label{23}
\end{equation}
Since the above equation does not have an exact solution for $\omega=-0.35$, we numerically find the roots of it. In Figure 8, we have plotted the roots of equation (\ref{23}) which are identified with $r_{1c}$, and $r_{2c}$. Also, in left panel of Figure 9, $r_{1c}$ and $r_{2c}$ have been plotted in one graph. The solid red curve corresponds to the 4$D$ EGB black holes in the absence of quintessence matter, $q=0$. The behaviour of $r_{1c}$, $r_{2c}$ and $r_+$ has been plotted in one graph, in the right panel of the figure and the photon radius are compared to the radius of the event horizon. According to this figure, we see that $r_{1c}>r_+$, while $r_{2c}<r_+$. Therefore, $r_{1c} \equiv r_c$ is the radius of the unstable circular orbit since it is always larger than the event horizon $r_+$. As is clear from the left panel of Figure 8, with increasing the GB coupling constant the photon radius, $r_{c}$, decreases while for a constant value of $\alpha$ by increasing the $q$ parameter the photon radius increases, which is in agreement with Table 1.

The numerical results for the radius of the photon sphere for 4$D$ EGB black hole with quintessence matter are compared to the case of 4$D$ EGB black hole, the Schwarzschild black hole and also the Schwarzschild black hole with quintessence matter, in Table 1. It shows that the presence of the positive GB coupling constant decreases the radius of the photon sphere of 4$D$ EGB black hole in comparison with the Schwarzschild black hole \cite{GB2}. This is due to the fact that the positive GB coupling $\alpha>0$, has the role of the dark energy, counteract the gravity and decreases the instability area around the black hole and thus the photon radius takes smaller values. However, the quintessence matter increases the radius of the photon sphere in comparison with 4$D$ EGB black holes as well as the Schwarzschild black hole. Also, we see that in the limit of $\alpha\rightarrow 0$ and $q=0$, $r_c=3M$ and $b_c=\sqrt{27}M$ which are respectively the radius of the photon sphere and impact parameter for unstable circular orbits of the Schwarzschild space-time \cite{Sch}.

The impact parameter at the radius of the photon sphere is also given by
\begin{equation}
\frac{L_c}{E_c}=\frac{r_c}{\sqrt{f(r_c)}}=\frac{r_c}{1+ \frac{r_c^2}{2 \alpha } \left[1-\sqrt{1+\frac{8 \alpha  M}{r_c^3}+\frac{8 \alpha  q}{ r_c^{1.95}}}\right]} \equiv b_c.
\label{24}
\end{equation}
The values of the impact parameter for different values of $q$ and $\alpha$ are presented in Table 1. As one can see, by increasing the GB coupling $\alpha$ the impact parameter decreases, while with increasing $q$ the values of impact parameter increase.

\begin{figure}[H]
\centering
\includegraphics[width=3.0in]{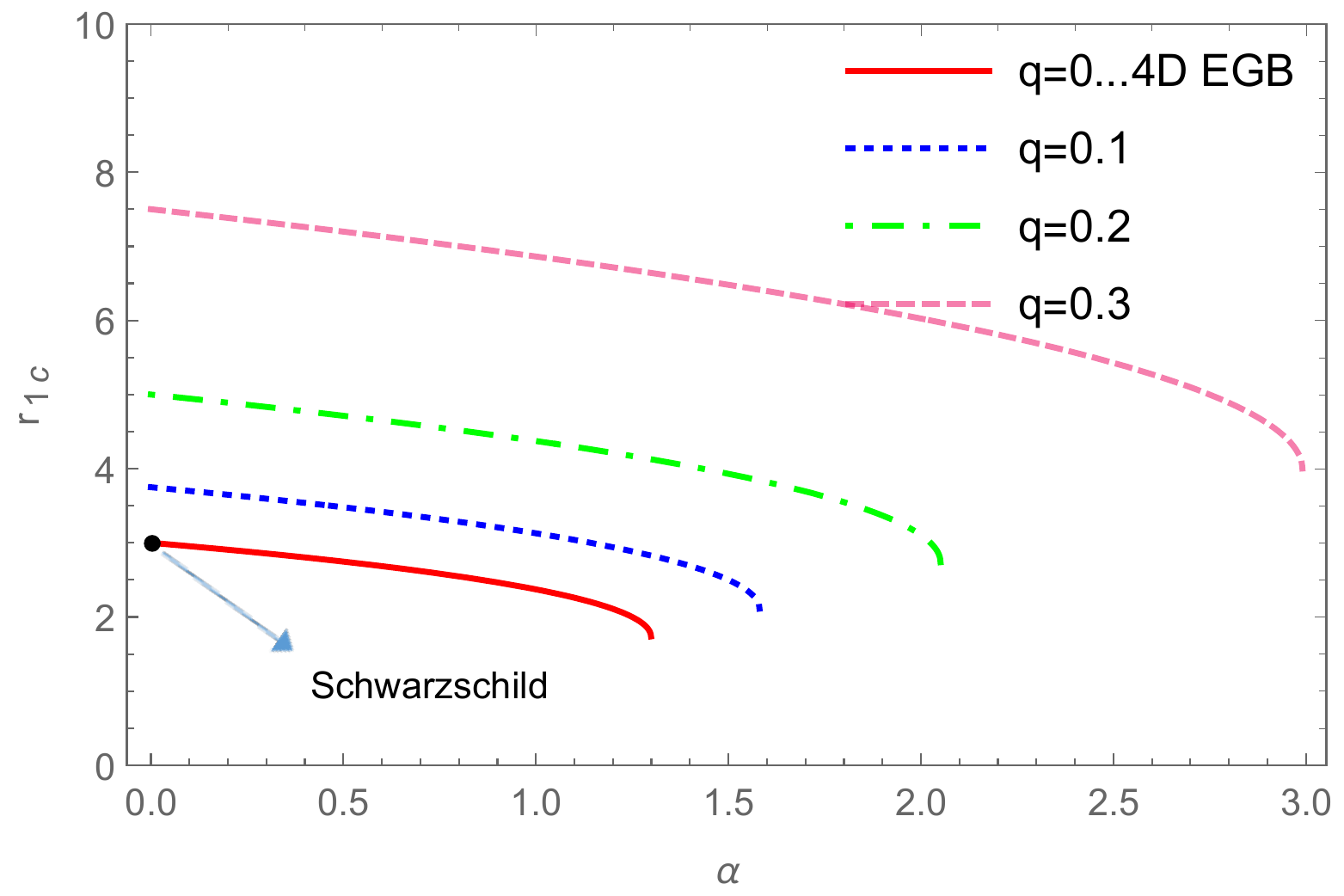}
\includegraphics[width=3.0in]{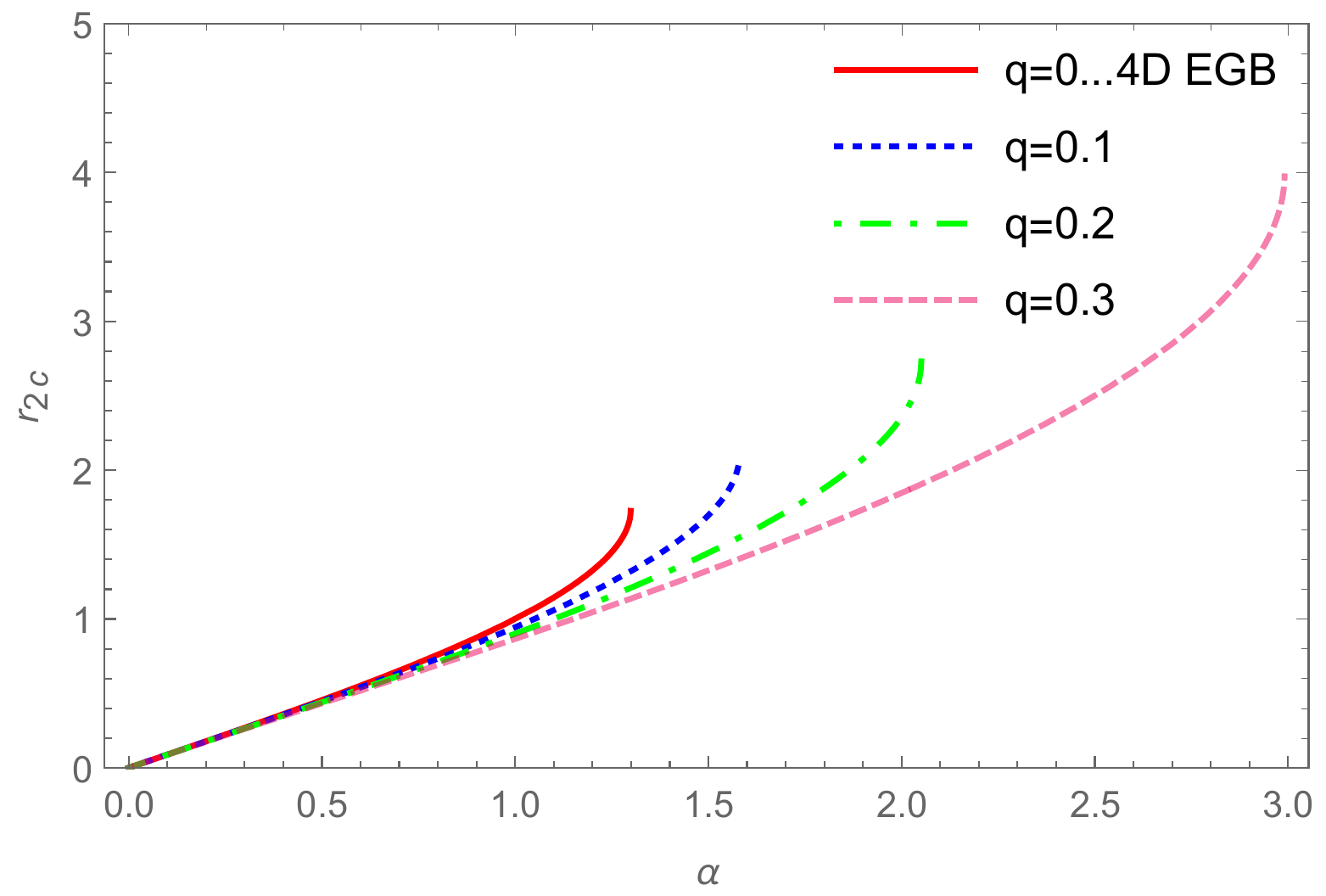}
\caption{Left panel: The photon radius $r_{1c}$ as a function of GB coupling constant for different values of $q$. Right panel: The photon radius $r_{2c}$ as a function of GB coupling constant for different values of $q$ and $L = 30$ with $M = 1$.}
\label{photonradiuas}
\end{figure}

\begin{figure}[H]
\centering
\includegraphics[width=3.0in]{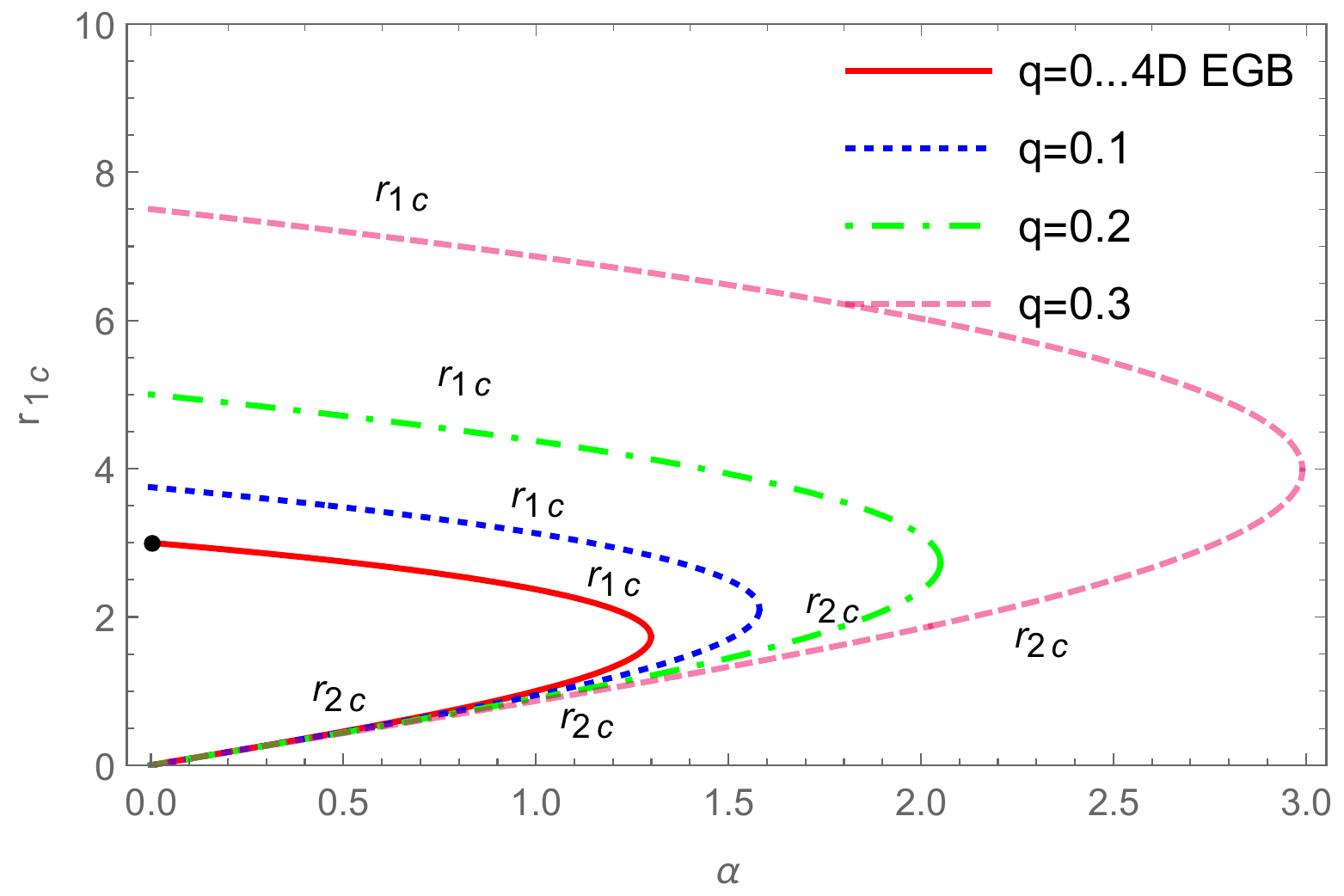}
\includegraphics[width=3.0in]{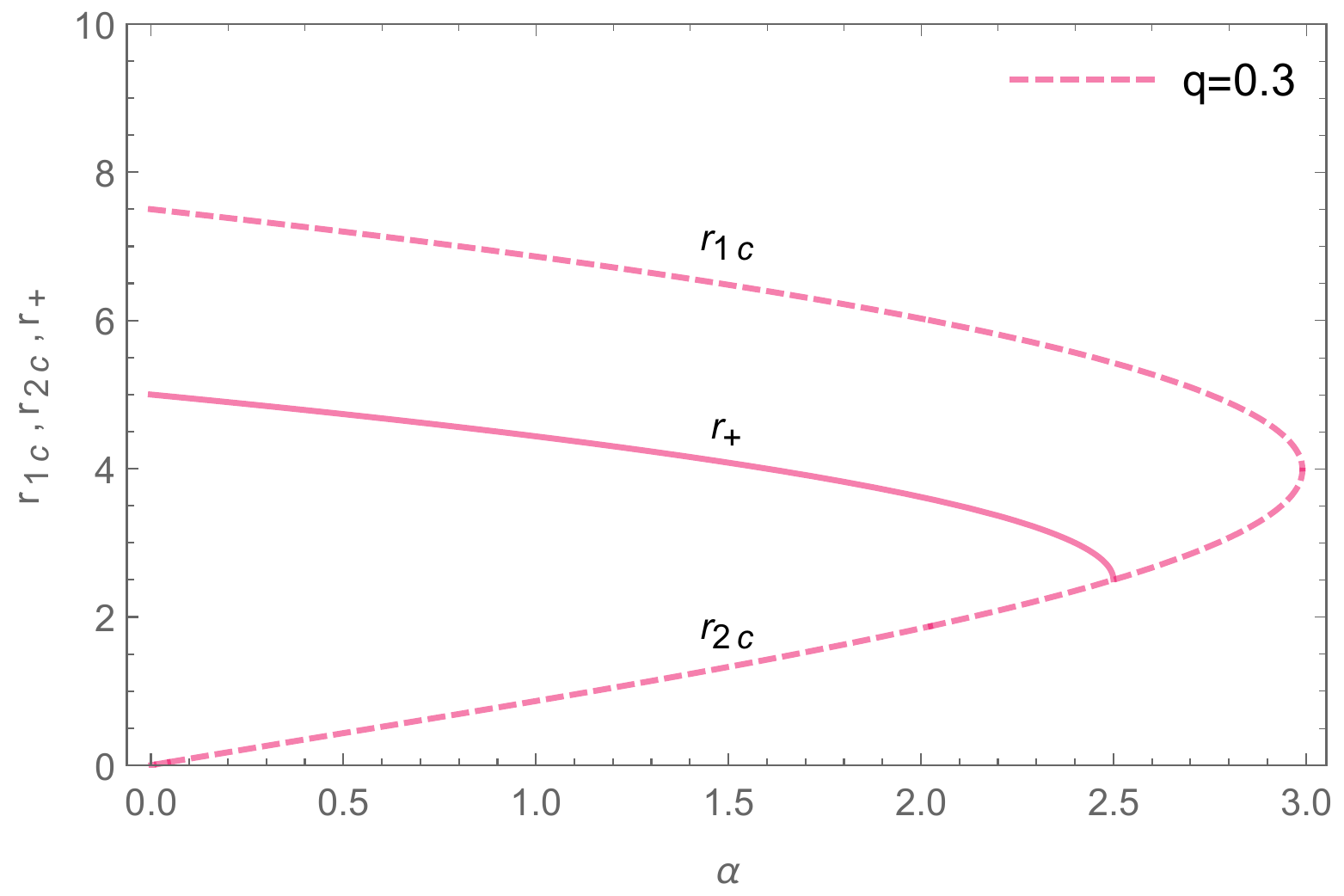}
\caption{Left panel: The photon radii $r_{1c}$ and $r_{2c}$ as a function of GB coupling constant for different values of $q$. Right panel: The comparison of the photon radii $r_{1c}$ and $r_{2c}$ with the event horizon $r_{+}$ for $q = 0.3$, $L = 30$ and $M = 1$.}
\label{photonradius2}
\end{figure}

\subsubsection{Unbounded orbits}
By substituting the function $f(r)$ from equation (\ref{5}) into the equation (\ref{13}), the equation of motion reads as
\begin{equation}
\frac{dr}{d\varphi} =\sqrt{\frac{r^4}{b^2}-r^2-\frac{r^4}{2\alpha}\left[1-\sqrt{1+\frac{8\alpha M}{r^3}+\frac{8\alpha q}{r^{1.95}}}\right]}.
\label{25}
\end{equation}
Since the above equation does not have an analytical solution, we numerically plot this equation for different values of energy, as we explained in the previous section. To this end, we first rewrite the above equation in terms of the new variable $u=\frac{1}{r}$ as follows
\begin{equation}
\frac{du}{d\varphi} = \sqrt{\frac{1}{b^2}-u^2-\frac{1}{2\alpha}\left[1-\sqrt{1+{8\alpha M u^3}+{8\alpha q u^{1.95}}}\right]}\equiv F(u).
\label{26}
\end{equation}
The geometry of the geodesics depends on the roots of the equation $F(u)=0$. To understand the behaviour of light rays in the space-time of 4$D$ EGB black holes surrounded by quintessence matter, we assume that light rays approach the black hole from the infinity. By considering the different values of the impact parameter in the above equation, we can display the different trajectories. For $b = b_c$ the light rays that approach to the black hole from $r > r_c$, revolve around the black hole at the radius of $r=r_c$. The plot of this unstable circular orbit is shown with orange curve in Figure 10 which relates to the light rays in region II in Figure 7. For $b > b_c$, the light rays approaching to the black hole from infinity, will be deflected at the turning point $u_0=\frac{1}{r_0}$ which is the root of the function $F(u)$, see Figure 7. This case corresponds to the unbounded null geodesics which is shown with blue curve in the Figure 11. However, in region III with $b < b_c$, the light rays approaching to the black hole from infinity cross the event horizon and enter the black hole. We have shown these falling orbits with the green curves in Figure 11.

\begin{figure}[H]
\centering
\includegraphics[width=3.0in]{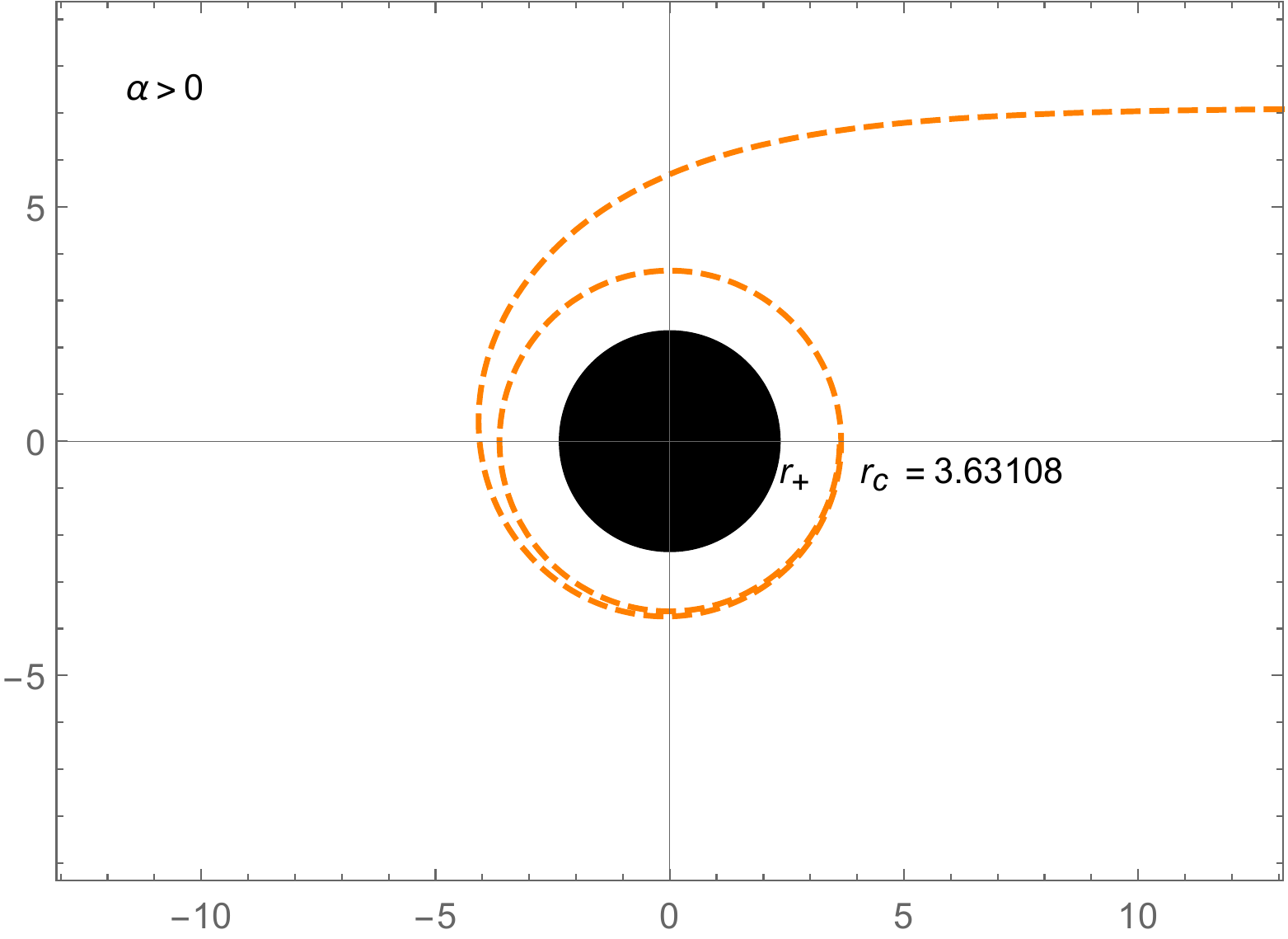}
\includegraphics[width=3.0in]{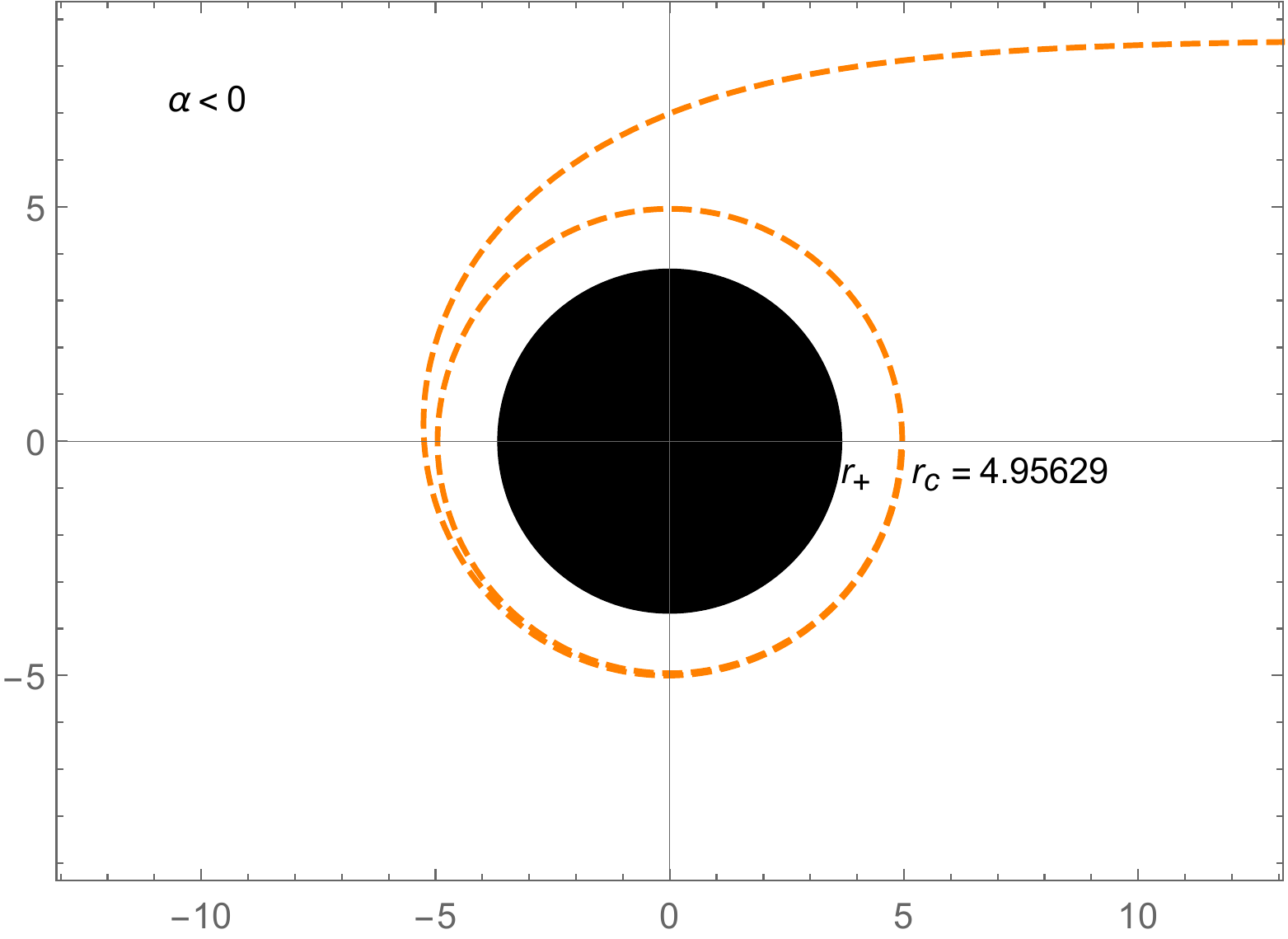}
\caption{The unstable circular orbit for the light ray approaches from $r > r_{c}$ to the 4$D$ EGB black hole surrounded by quintessence matter. Left panel: for positive GB coupling $\alpha = 0.3$, $q  =0.1$. Right panel: for negative GB coupling $\alpha = -3.3$, $q  =0.1$ and $M = 1$. In each panel the event horizon is at $r_{+} =  2.36768$ and $r_{+} = 3.68224$, respectively.}
\label{circular}
\end{figure}

\begin{figure}[H]
\centering
\includegraphics[width=3.0in]{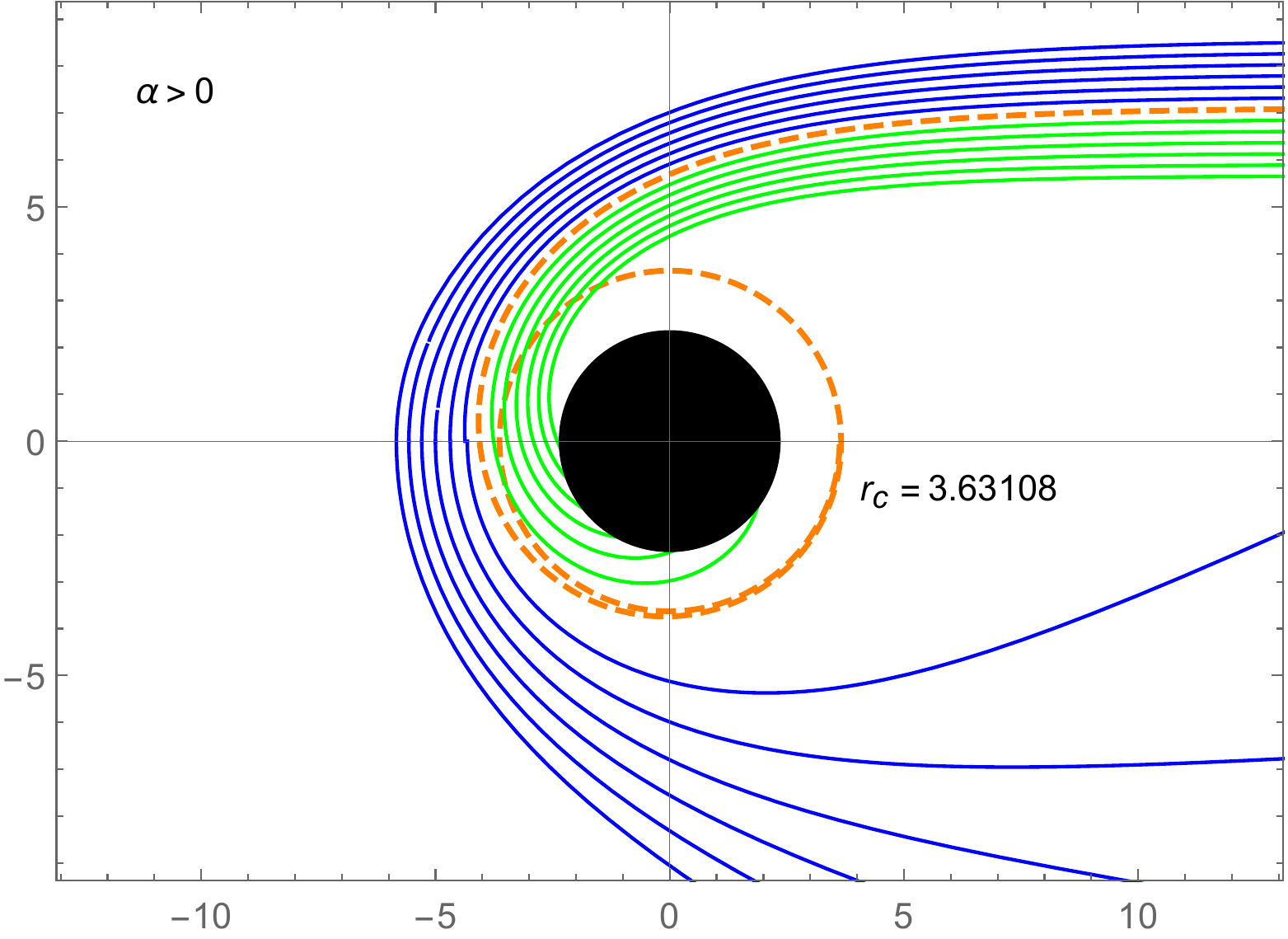}
\includegraphics[width=3.0in]{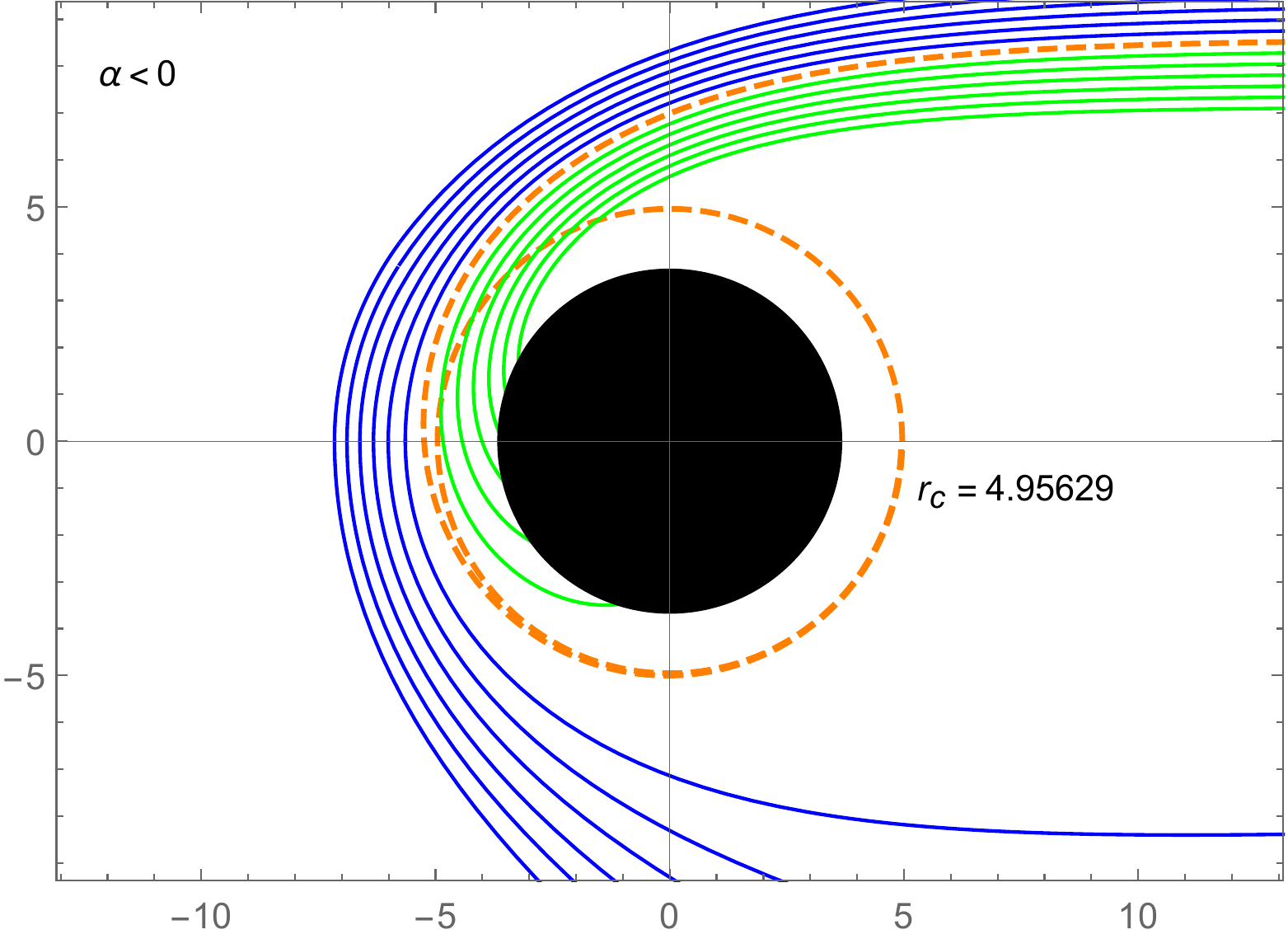}
\caption{The trajectory of the photons in polar coordinate $(r,\varphi)$. Left panel: for positive GB coupling $\alpha = 0.3$, $q  =0.1$. Right panel: for negative GB coupling $\alpha = -3.3$, $q  =0.1$ and $M = 1$. The orange dashed curve shows the unstable circular orbit, $b = b_c$, the green solid curves correspond to the null geodesics with $b < b_{c}$ and the blue curves correspond to $b > b{c}$. For all the null geodesics, we considered the spacing between impact parameters as $\Delta b=0.25$. In each panel the solid disk shows the black hole with the event horizon at $r_{+} =  2.36768$ and $r_{+} = 3.68224$, respectively.}
\label{polar-plot}
\end{figure}

\section{Shadow of quintessence 4$D$ EGB black holes}
In order to determine the shape and the size of the black hole shadow one need to obtain two celestial coordinates $X$ and $Y$ as follows \cite{Sch}
\begin{equation}
X=\lim_{r_0\rightarrow\infty}\left(-r_0^2\sin\theta_0\frac{d\varphi}{dr}\right),
\label{n1}
\end{equation}
\begin{equation}
Y=\lim_{r_0\rightarrow\infty}\left(r_0^2\frac{d\theta}{dr}\right),
\label{n2}
\end{equation}
where $r_0$ is the distance from the black hole to the observer and $\theta_0$ is the inclination angle between the line of sight of the observer and the rotation axis of the black hole. Using the geodesic equations and substituting the expressions $\frac{d\varphi}{dr}$ and $\frac{d\theta}{dr}$ into the above equations one can find the equations which relate the celestial coordinates to the constants of motion. However, in the case of our study where the observer is situated in the equatorial plane with the inclination angle $\theta_0=\frac{\pi}{2}$, it is easy to see that the radius of the shadow is equivalent to the critical impact parameter given by the following equation
\begin{equation}
R_s\equiv\sqrt{X^2+Y^2}=b_c.
\label{n3}
\end{equation}
In Figure 12, we have displayed the boundary of the shadow of quintessence 4$D$ EGB black holes for different values of quintessence parameter $q$ and the GB coupling $\alpha$. In the top panels of figure we have plotted the shadow of 4$D$ EGB black holes without quintessence, $q=0$. As is clear, for positive values of the GB coupling $\alpha$ the shadow size of EGB black holes is smaller than that for the Schwarzschild black hole, while these are larger for negative $\alpha$. The size of the shadow for quintessence EGB black holes with $q=0.1$ is shown in the bottom panels. We see that for positive $\alpha$ the shadow radius of quintessence EGB black holes is smaller than that of the quintessence Schwarzschild black hole, while it is larger for the negative ones. Moreover, it is easy to see that the shadow size of quintessence 4$D$ EGB black holes is larger than that of 4$D$ EGB black holes without quintessence; namely the presence of the quintessence matter increases the shadow radius of the black hole which is in agreement with Table 1.

\begin{figure}[H]
\centering
\includegraphics[width=2.25in]{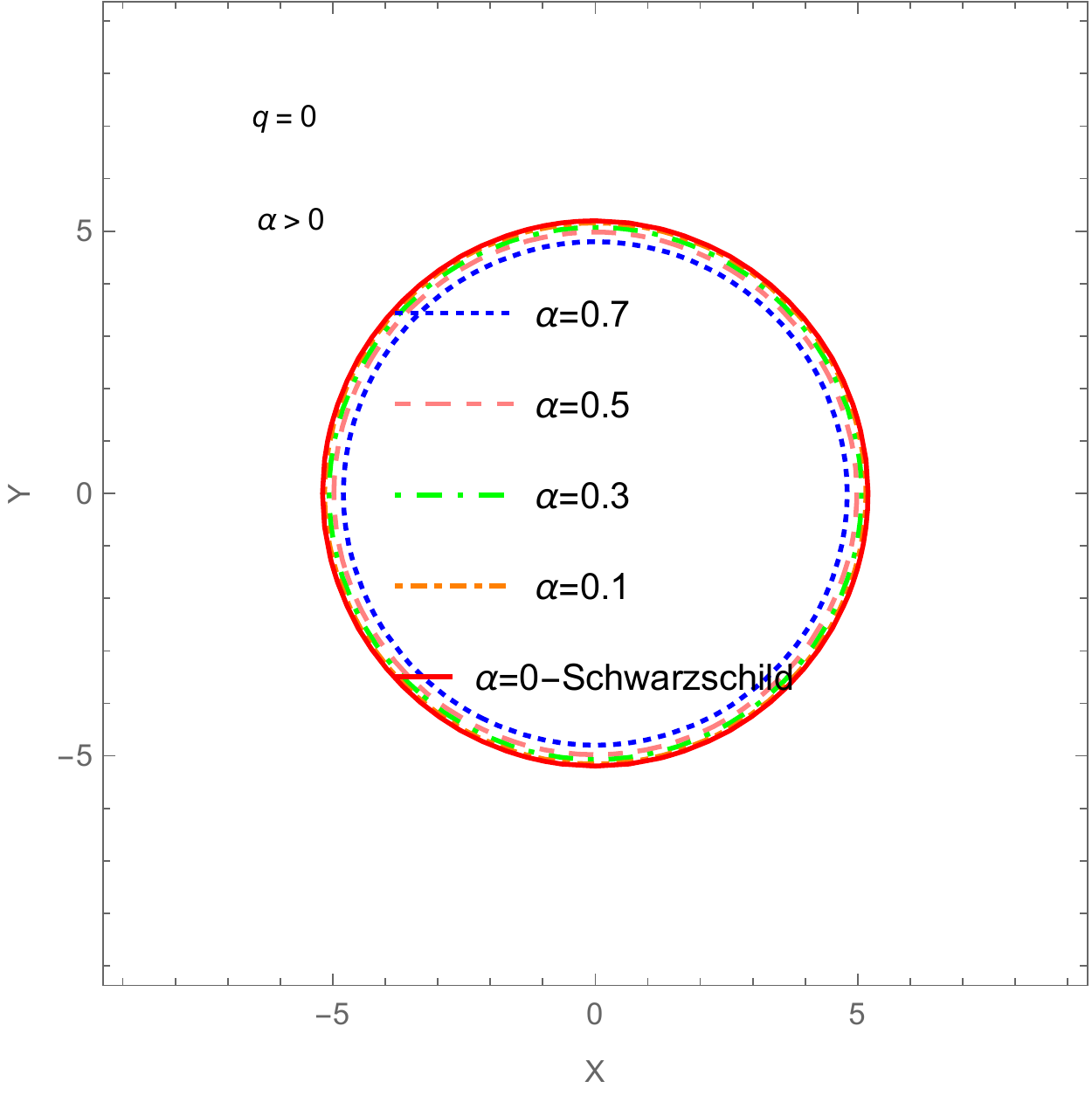}
\includegraphics[width=2.25in]{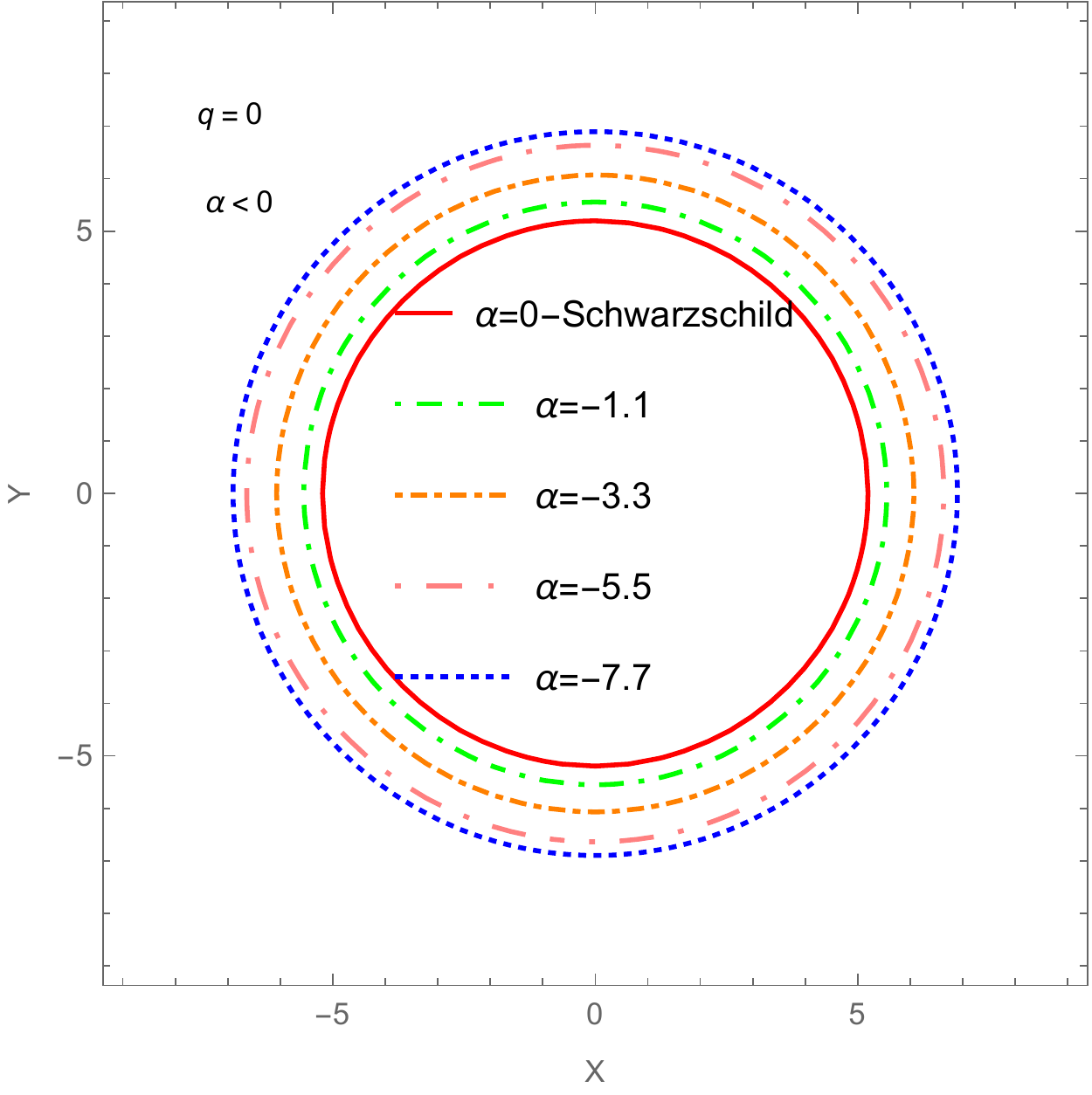}
\includegraphics[width=2.25in]{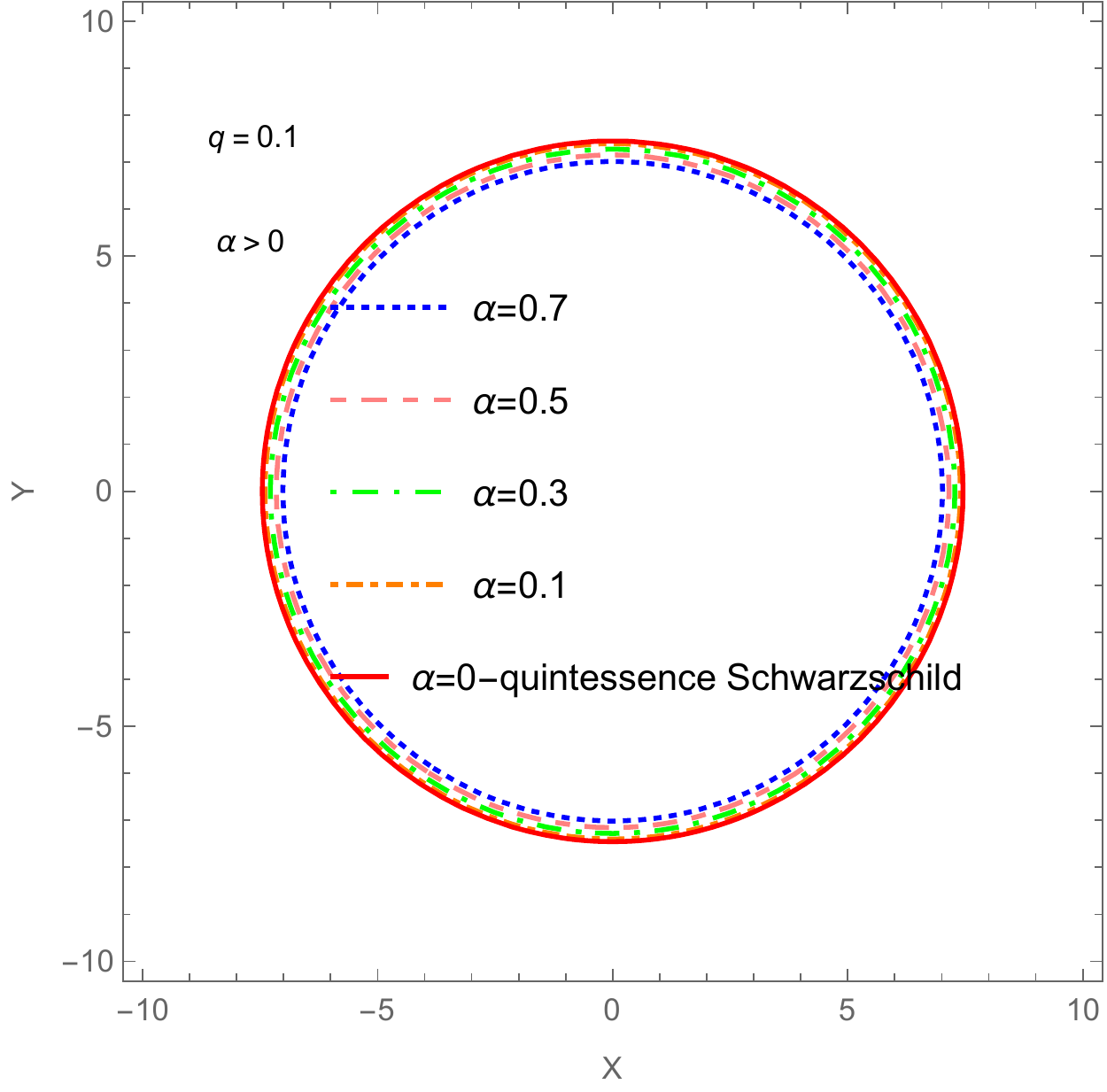}
\includegraphics[width=2.25in]{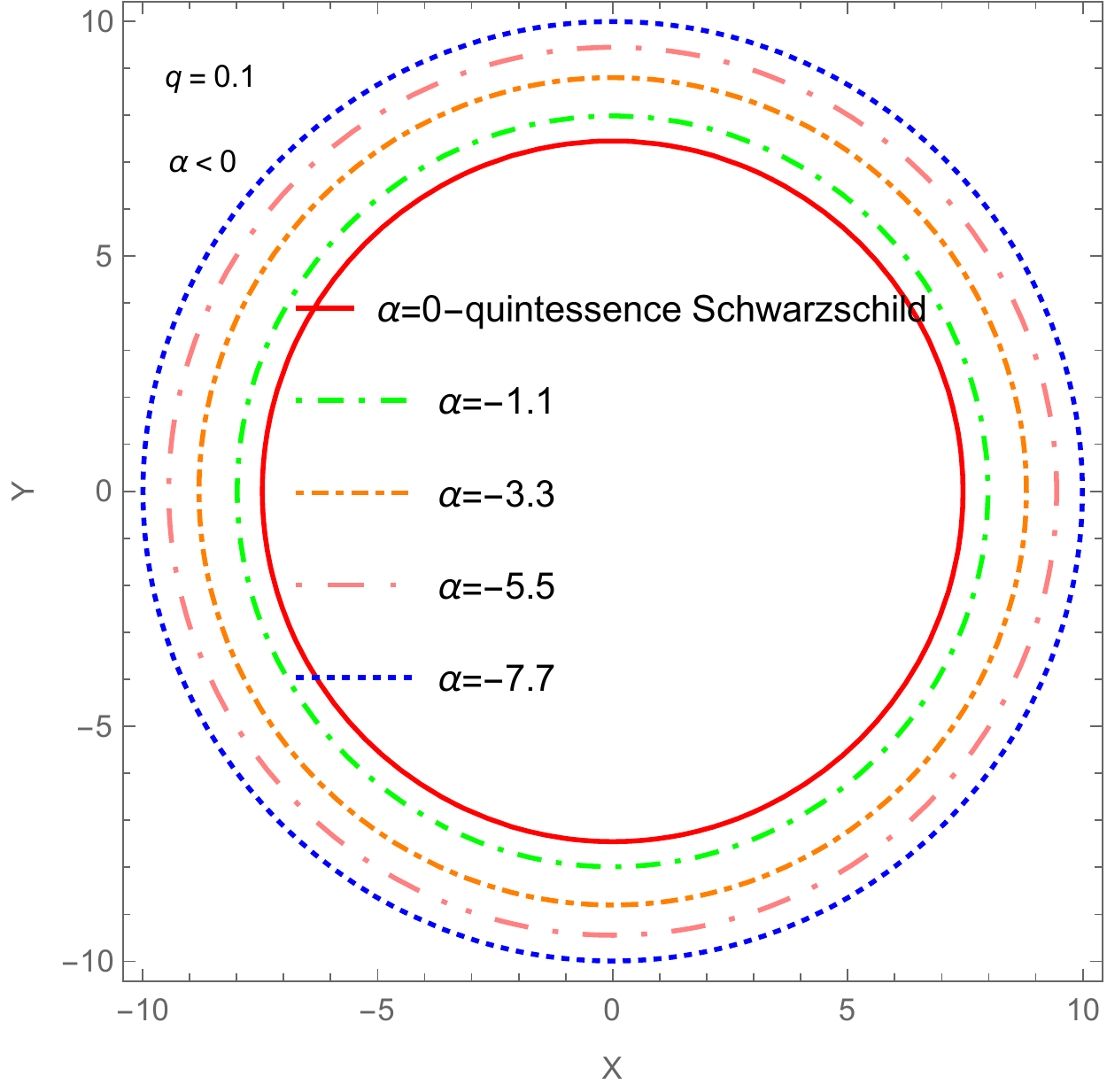}
\caption{Top-left panel and top-right panel the shadow cast by 4$D$ EGB black holes without quintessence. Bottom-left panel and bottom-right panel the shadow cast by 4$D$ EGB black holes with quintessence.}
\label{shadow}
\end{figure}

\begin{figure}[H]
\centering
\includegraphics[width=3.0in]{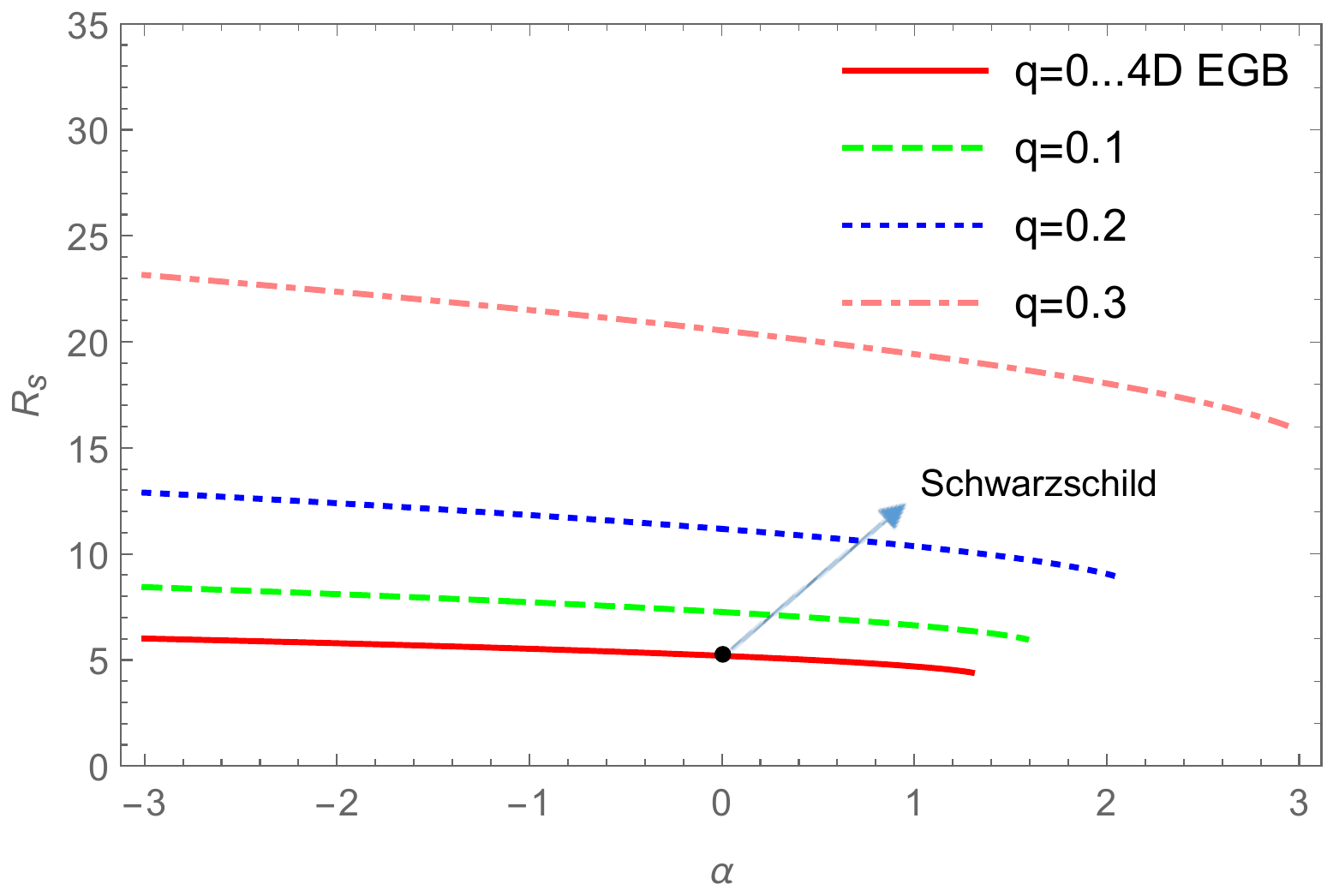}
\includegraphics[width=3.0in]{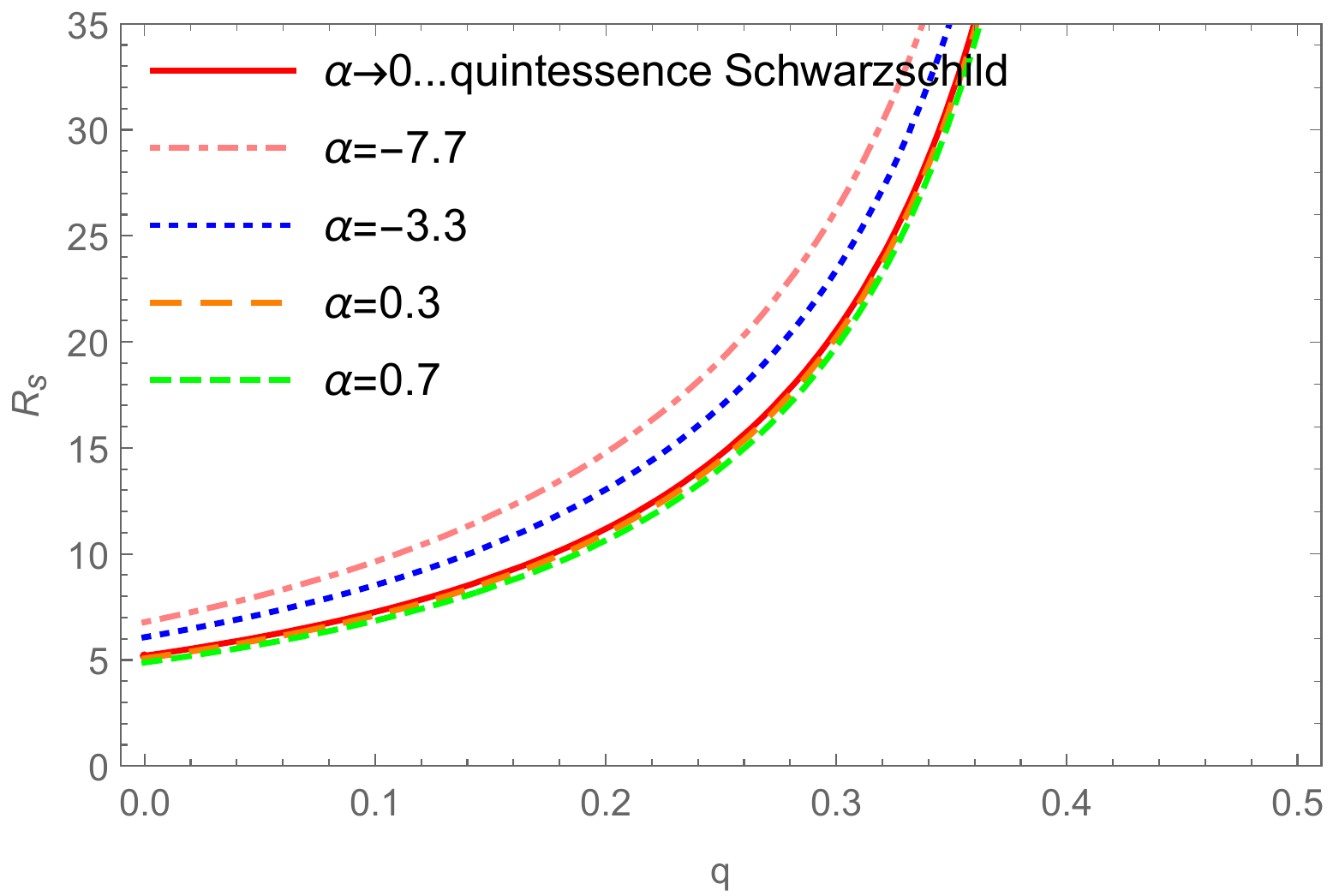}
\caption{ The radius of shadow $R_s$. Left panel: for different values of $q$ as a function of GB parameter $\alpha$. Right panel: for different values of the GB coupling $\alpha$ as a function of $q$. The solid curves correspond to 4$D$ EGB black holes without quintessence matter and quintessence Schwarzschild black hole in the left and right panels, respectively.}
\label{delta}
\end{figure}

Also, in the left panel of Figure 13 we have plotted the radius of shadow for 4$D$ EGB black holes surrounded by quintessence matter for different values of quintessence parameter $q$. The solid red curve corresponds to 4$D$ EGB black holes without quintessence, $q=0$. We see that for a fixed value of $\alpha$ by increasing the $q$ parameter the shadow radius increases. In the case of $\alpha=0$ and $q=0$ we retain the radius of shadow for the Schwarzschild black hole, i.e, $R_s=3\sqrt{3}$, the solid point on the red curve. The right panel of the figure shows the radius of shadow for different values of the GB coupling $\alpha$. The solid red curve corresponds to the quintessence Schwarzschild black holes in the limit of $\alpha\rightarrow 0$. As can be seen, for a fixed value of $q$ by increasing the $\alpha$ parameter the shadow size decreases. Also, it is clear that the shadow radius for $\alpha < 0$ is larger than the quintessence Schwarzschild black hole, while it is smaller for $\alpha > 0$.

\section{Deflection angle by quintessence 4$D$ EGB black holes}
The problem of light bending in the space-time of 4$D$ EGB black holes has been considered in \cite{lensing3}. Also, the deflection angle of charged massive particles has been calculated in \cite{Charge}. Now, in this section we are going to calculate the bending angle of light for 4$D$ EGB black holes surrounded by quintessence matter for $\omega_{q}=-0.35$. As we mentioned in the previous section, when the light rays with $b>b_c$ approach to the black hole, will be deflected at the turning point $r=r_0$, see Figure 7. So, the first step is to find the closest approach $r_0$. To this end, we need to solve the equation $\frac{dr}{d\varphi}=0$. From equation (\ref{25}) we find
\begin{equation}
\left(\frac{1}{r^2}\frac{d\varphi}{dr}\right)^2 = {\frac{1}{b^2}-\frac{1}{r^2}-\frac{1}{2\alpha}\left[1-\sqrt{1+\frac{8\alpha M}{r^3}+\frac{8\alpha q}{r^{1.95}}}\right]} \equiv \xi(r).
\label{27}
\end{equation}
As is clear, the value of impact parameter has an important role in determining the closest approach. In Figure 14, we have displayed $\xi(r)$ as a function of $r$ for different values of the impact parameter. The figure shows that the function $\xi(r)$ has two roots at $r=r_0$ and $r=r_1$, for $b>b_c$. Also, according to the Figure 7 we see that $r_0>r_c$. Therefore, the root at $r=r_0$ is considered as the closest approach. Moreover, according to the figure, by increasing the impact parameter $b$ the closest approach $r_0$ also increases.

Using equation (\ref{26}) the bending angle is given by

\begin{equation}
\delta = 2\int_{0}^{u_0}\bigg|\frac{d\varphi}{du}\bigg|du-\pi=2\int_{0}^{u_0}\frac{1}{\sqrt{\frac{1}{b^2}-u^2-\frac{1}{2\alpha}\left[1-\sqrt{1+{8\alpha M u^3}+{8\alpha q u^{1.95}}}\right]}}du-\pi.
\label{28}
\end{equation}
In the right panel of Figure 15, the bending angle as a function of $u_0$ is plotted for $q=0.1$ and different values of the GB coupling. It is easy to see that in the limit $u_0\rightarrow 0$, the bending angle tends to zero, as one would expect. As one decreases the $\alpha$ parameter, the effect of the GB term as a candidate for dark energy also decreases, and thus the strength of gravitational attraction increases leads to an increase of the bending angle. Therefore, in the case of $ \alpha\rightarrow 0$ which corresponds to the Schwarzschild black hole surrounded by quintessence matter, we have the largest values of the deflection angle. In order to see the effect of the $q$ parameter on the light bending, we have shown the bending angle as function of $u_0$  for $\alpha=0.7$ and different values of $q$ in the left panel of Figure 15. According to the figure, with decreasing $q$ the bending angle also decreases, so that for $q\rightarrow0$ which corresponds to the 4$D$ EGB black holes, the bending angle has the minimum value.

\begin{figure}[H]
\centering
\includegraphics[width=3.0in]{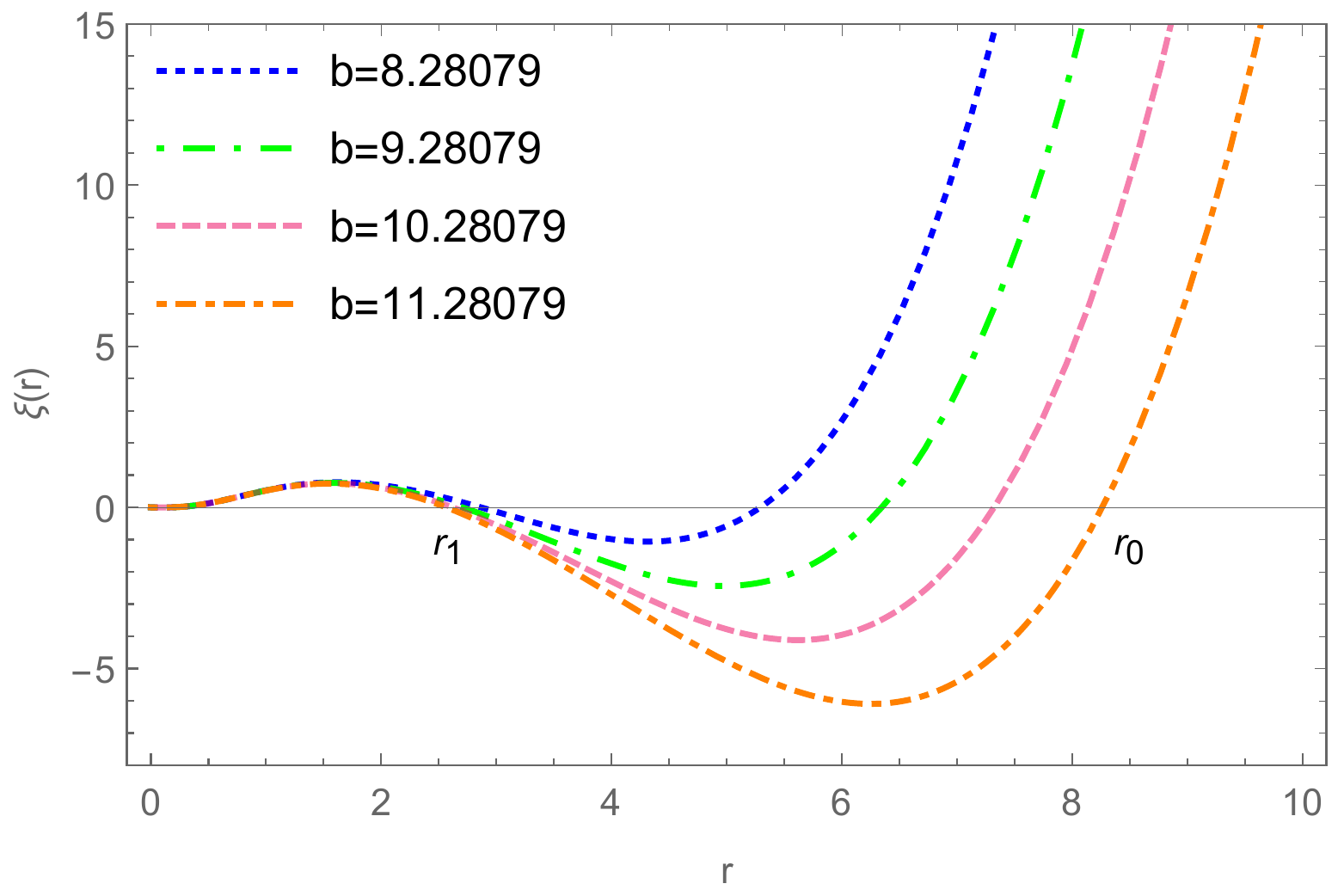}
\caption{The behavior of $\xi$ as a function of $r$ for different values of impact parameter $b$ for $\alpha = 0.3$, $q = 0.1$ and $M = 1$. }
\label{distance}
\end{figure}

\begin{figure}[H]
\centering
\includegraphics[width=3.0in]{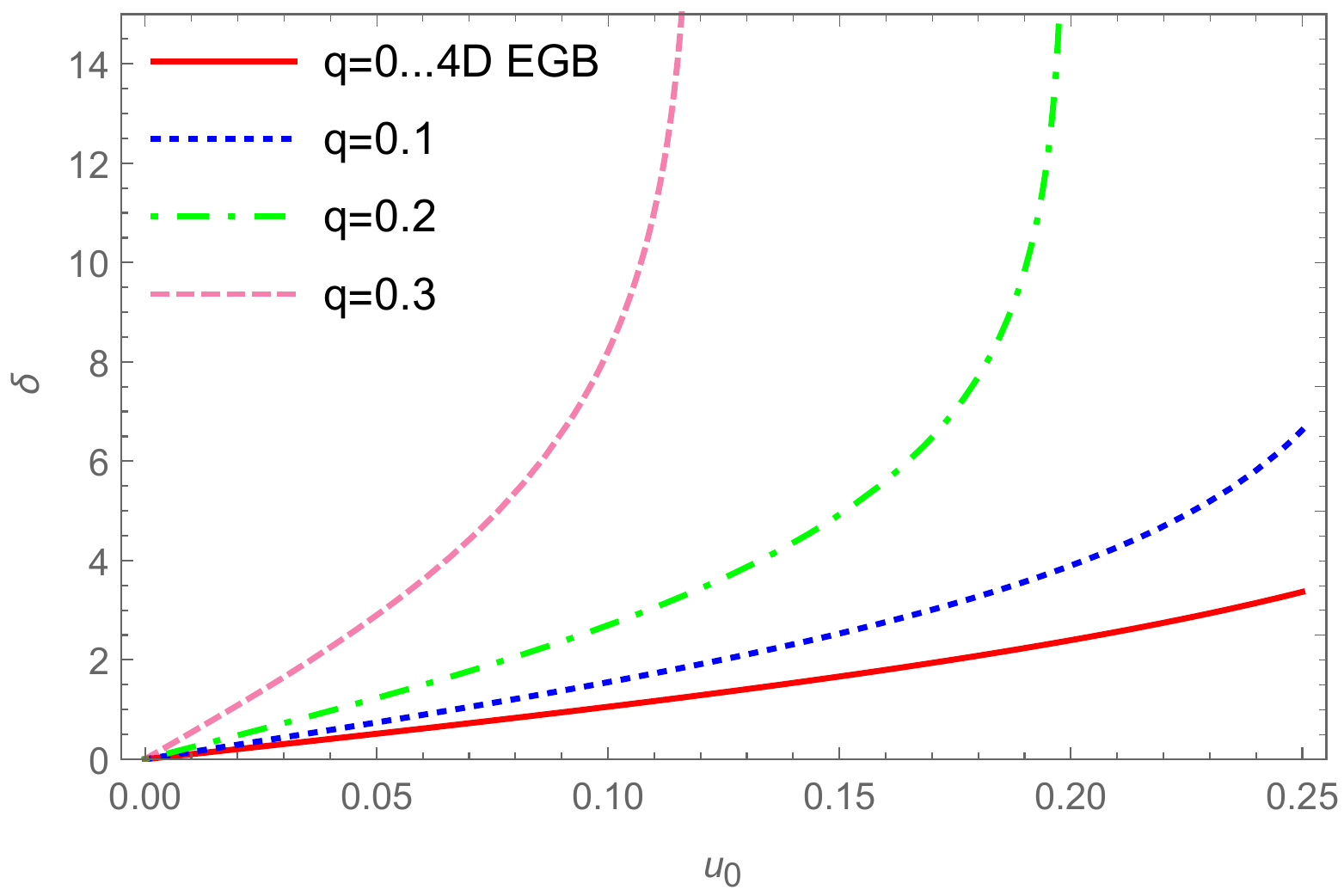}
\includegraphics[width=3.0in]{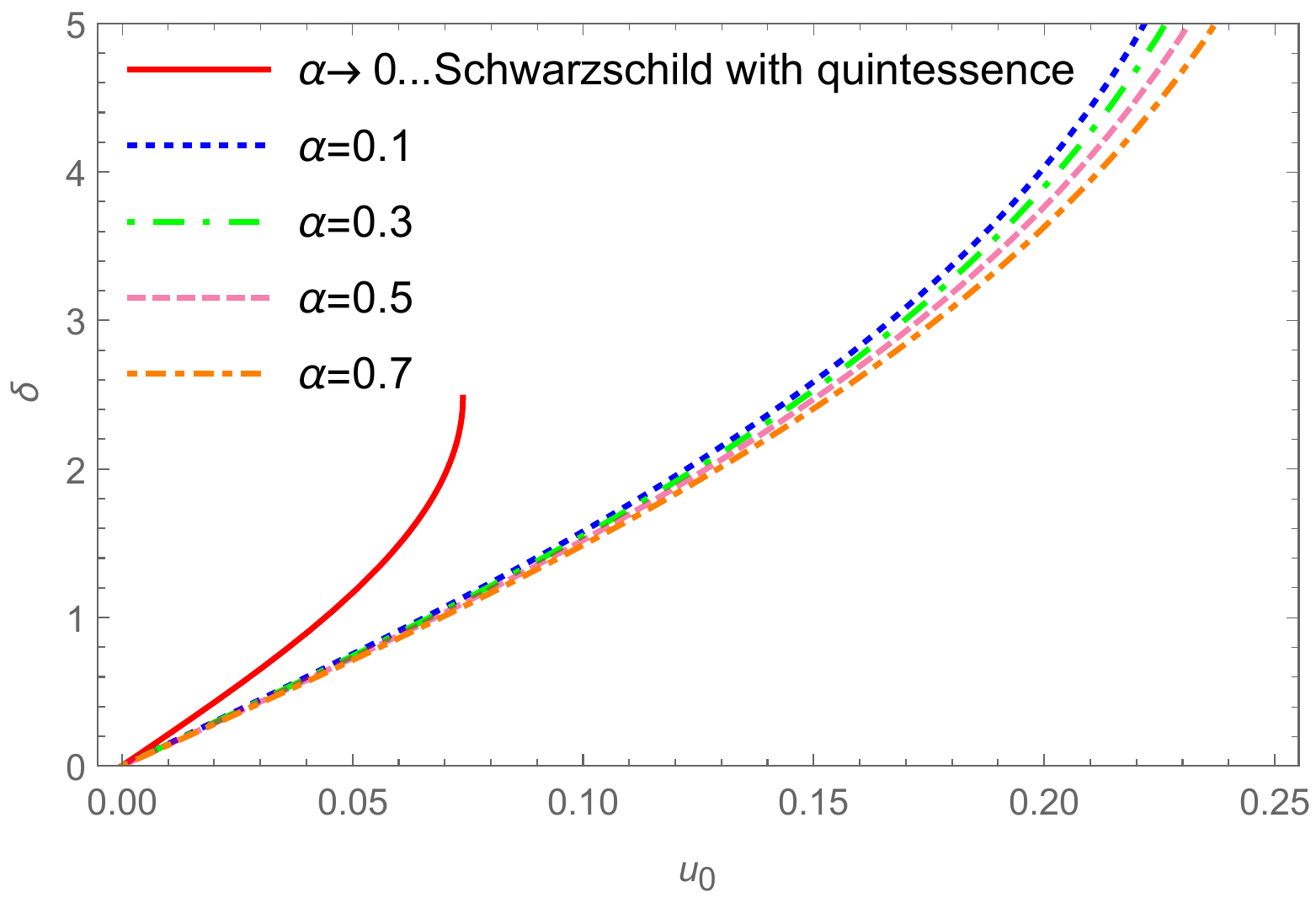}
\caption{The bending angle versus the inverse of closet approach $u_0=\frac{1}{r_0}$. Left panel: for $\alpha = 0.3$ and different values of $q$. Right panel: for $q = 0.1$ and different values of the GB coupling $\alpha$.}
\label{delta}
\end{figure}

\section{Conclusions}
The exact $4D$ static spherically symmetric black hole solutions to EGB gravity surrounded by quintessence matter have been derived in Ref. \cite{QEGB}. In this paper, we considered these 4$D$ EGB black holes with quintessence matter and studied the shadow and null geodesics around them and investigated the effects of both the GB coupling constant and the quintessence parameter on the motion of light rays.

It is shown that depending on the value of $\omega_{q}$, the metric of quintessence 4$D$ EGB black holes could have one, two or three horizons. So, we first presented the structure of the black hole event horizon for $\omega_{q}=-0.35$ and then for this special case we studied both radial and non-radial null geodesics in details. For non-radial geodesics, we have numerically obtained the physical quantities such as the effective potential, radius of the photon sphere and impact parameter for different values of the GB coupling and the quintessence parameter, which the results summarized in Table 1. We found that for a given value of $q$, with increasing the GB coupling the values of the event horizon and the photon radius decrease. This result is to be expected, because a positive GB coupling plays the role of dark energy, counteract gravity and thus the event horizon and photon radius take smaller values. However, it is shown that for a fixed value of $\alpha$, by increasing the quintessence parameter $q$, both the event horizon and the photon radius also increase. Therefore, the presence of the quintessence matter around a 4$D$ EGB black hole increases the values of the event horizon and the photon sphere. The achieved results are compared to that obtained for 4$D$ EGB black holes without quintessence \cite{GB2} and to the Schwarzschild black hole surrounded by quintessence matter \cite{Qu2}.

Also, we have analyzed the null trajectories around quintessence 4$D$ EGB black holes for different values of the impact parameter and we have numerically plotted the unstable circular and unbounded orbits of photons in details. Furthermore, we have studied the role of the GB coupling and the quintessence parameter on the shadow size of quintessence 4$D$ EGB black holes and found that the shape of the shadow is a perfect circle which its radius increases with the quintessence parameter. However, for a given value of $q$, the shadow radius decreases for positive GB coupling and increases for negative ones. Finally, as an physical application of null geodesics we calculated the deflection angle around 4$D$ EGB black holes with quintessence and studied the effects of the GB coupling and the quintessence parameter on it. The study of null geodesics of charged black holes in 4$D$ EGB  gravity surrounded by quintessence matter is the subject of the future investigations.

\end{document}